\documentclass[a4paper]{aa}
\usepackage{natbib}
\usepackage{xspace}
\usepackage{amssymb}
\usepackage{amsmath}
\usepackage{graphicx}
\usepackage{rotating}
\usepackage{natbib}

\def\aap{A\& A}

\def\mnras{MNRAS}

\def\apj{ApJ}

\def\apjl{ApJL}

\def\cs{\ensuremath{c_\mathrm{s}}\xspace}
\def\teddy{\ensuremath{\tau_\mathrm{ed}}\xspace}
\def\omeddy{\ensuremath{\omega_\mathrm{ed}}\xspace}
\def\tsett{\ensuremath{\tau_\mathrm{sett}}\xspace}
\def\tdrift{\ensuremath{\tau_\mathrm{drift}}\xspace}
\def\tage{\ensuremath{\tau_\mathrm{age}}\xspace}

\def\Tgas{\ensuremath{T_\mathrm{gas}}\xspace}

\def\comma{\,,}
\def\fullstop{\,.}

\begin{document}
  
  \title{Survival of the mm-cm size grain population observed in
  protoplanetary disks}

  \titlerunning{Radial drift of mm-cm size grains}
  
  \authorrunning{Brauer,
    Dullemond,
    Johansen,
    Henning,
    Klahr \and 
    Natta}
  
  \author{F.~Brauer \inst{1} \and 
    C.P.~Dullemond \inst{1} \and 
    A.~Johansen \inst{1} \and 
    Th.~Henning \inst{1} \and 
    H.~Klahr \inst{1} \and
    A.~Natta \inst{2}}
  
  \date{\today} 

  \institute{Max-Planck-Institut f\"ur Astronomie, K\"onigstuhl 17, D--69117
  Heidelberg, Germany; e--mail: brauer@mpia.de \and Osservatorio Astrofisico
  di Arcetri, Largo Enrico Fermi 5, 50125 Firenze, Italy; e--mail:
  natta@arcetri.astro.it}
  
  \abstract{Millimeter interferometry provides evidence for the presence of mm
  to cm size `pebbles' in the outer parts of disks around pre-main-sequence
  stars. The observations suggest that large grains are produced relatively
  early in disk evolution ($<$ 1 Myr) and remain at large radii for longer
  periods of time (5 to 10 Myr). Simple theoretical estimates of the radial
  drift time of solid particles, however, imply that they would drift inward
  over a time scale of less than 0.1 Myr. In this paper, we address this
  conflict between theory and observation, using more detailed theoretical
  models, including the effects of sedimentation, collective drag forces and
  turbulent viscosity. We find that, although these effects slow down the
  radial drift of the dust particles, this reduction is not sufficient to
  explain the observationally determined long survival time of mm/cm-sized
  grains in protoplanetary disks. However, if for some reason the gas to dust
  ratio in the disk is reduced by at least a factor of 20 from the canonical
  value of 100 (for instance through photoevaporation of the gas), then the
  radial drift time scales become sufficiently large to be in agreement with
  observations.}

\maketitle


\section{Introduction}
Millimeter and sub-millimeter observations have shown the presence of large
amounts of millimeter to centimeter-sized grains in the outer regions ($\sim$
100 AU) of disks around Herbig Ae and T Tauri stars
\citep{Tes03,Rod06,Wil05,natta06}. The presence of these grains, which are
much larger than the grains typically found in the interstellar medium, is
often regarded as evidence that the first steps of planet formation are taking
place in these disks. The presence of such large grains, however, also poses a
major problem. According to simple theoretical considerations, grains of this
size undergo a rapid radial drift \citep{Whipple72,Wei77}, which causes them
to disappear from the outer disk in a very short time
\citep{TakLin05,klabod06,alearm07}. However the typical age of protoplanetary
disks that are observed at millimeter wavelengths is a few Million years,
which is much longer than this radial drift time scale. Takeuchi \& Lin
propose that either the entire grain growth process is slow or that the grains
are the collision products of a population of even larger bodies ($\gtrsim$ 10
m). The second explanation requires that in addition to the grain population
that is observed at mm wavelengths, there is a population of smaller/larger
bodies which act as a reservoir of solid material from which mm/cm-sized
grains are continuously produced. The problem is that if the drift time scale
is, for example, 20 times shorter than the disk life time, this reservoir of
larger bodies must contain at least 20 times more mass than the observed dust
mass.  If it is assumed that the particle size distribution follows a powerlaw
then the total mass of the disk and the minimum upper particle size of this
distribution can directly be calculated from the slope of the mm flux of the
protoplanetary disk.  This analysis shows that the amount of dust responsible
for the millimeter fluxes of these disks is in many cases already very high,
of the order of $10^{-3}M_{\odot}$ or even higher
\citep{Tes03,nattesner04,Wil05,Rod06,rodphd}. A 20 times more massive
reservoir of larger (non-observable) bodies is then clearly unrealistic. These
arguments suggest that perhaps the standard theoretical estimate of the radial
drift may be not applicable.

Interestingly, a related drift problem is present for the theory of
planetesimal formation. The drift problem for mm/cm-sized particles at 100 AU,
that will be investigated in this paper, similarly shows up for meter-sized
particles at 1 AU. The radial drift of these large bodies in the inner parts
of the disk is so high that they should drift into the evaporation zone over
time scales of $\sim 10^2$ yrs. This is one of the fundamental and unresolved
problems of planet formation \citep{domppv07}. In that sense, the cm problem
at 100 AU is a proxy for the meter problem at 1 AU, and figuring out a
solution at 100 AU may give important clues to what happens at 1 AU.

The goal of this paper is to study the radial drift of mm/cm-sized particles
in more detail. We will investigate the magnitude of the drift problem, and
which effects might keep the grains for a few Myrs in the outer parts of the
disk. To address this issue we will proceed in steps. In the first step we
will review the radial drift of individual particles
\citep{Whipple72,Wei77}. This section will show that the drift time scale of
such particles is orders of magnitude smaller than the age of the disks
observed (5 to 10 Myrs). In a second step, we explore the possibility that
collective effects of the dust might slow down the drift. Collective effects
take place when the dust settles into a thin midplane layer
\citep{DubMorSte95,GarFouLin04,schr04}. This process increases locally the
dust-to-gas ratio, and the dynamics of the dust starts to affect the gas
motion \citep{NakSekHay86,jhk06}. This may, in turn, reduce the relative
velocities between the dust and the gas, and hence reduce the head wind that
causes radial drift. We will investigate the magnitude of the reduction and if
thin midplane layers yield a possibility to increase the drift time scales to
some Myrs. In the third step we improve on these calculations by including
vertical angular momentum exchanges in the disk through turbulent
viscosity. Finally, we speculate on other potential ways in which mm/cm-sized
grains could be prevented from drifting inward on a time scale shorter than
the life time of the disk: particle trapping in vortices and gas pressure
maxima \citep{barsom95,KlaHen97,fronel05,JohKlaHen06}, spiral arms
\citep{Rice04} and photoevaporation of the gas leaving the dust behind
\citep{acp06}.

Over the last decades various papers determined the radial drift of dust
particles and the structure and dynamics of thin midplane dust layers. In
particular the latter problem has attracted much attention, but for an
entirely different reason than ours: Gravitational instabilities in thin
midplane dust layers are thought to be a possible origin of planetesimals
\citep{GolWar73}. A lively debate has since appeared about the viability of
this concept, spurring various papers including models of midplane dust layers
\citep{Wei80,sek98,Wei06,dg,YouShu02,YouChi04,jhk06}. The richness
of this literature gives an indication of the complexity of the
problem. Hence, due to this complexity only a sub-set of the possible physical
effects are considered in these dust layer models. In particular the
collective effects of the dust and the effects of vertical and radial
viscosity have not been studied yet in combination. Therefore, another goal of
this paper is to present a model of dense dust midplane layers that include a
multitude of physical effects, albeit still in the form of a 1-D vertical
model.

In this paper we do not consider dust particle coagulation. Larger particles
in protostellar disks beyond sizes that can be found in the interstellar
medium form by collisional sticking due to relative motions of the dust
\citep{beck00,BluWur99}.  However, at 100~AU in the disk the dust particles
rapidly drift inwards before they can even grow to sizes that are discussed
this paper (Brauer et al. in prep.).  In the present paper we will ignore the
issue of the formation of these mm/cm size grains and assume that all the dust
is in the form of grains of a given size (a 'monodisperse' size
distribution). We will then investigate whether these grains can remain in the
outer regions of the disk for a sufficiently long time.

The structure of this paper is as follows. In section~\ref{rdotd} we discuss
the radial drift of the dust and its radial drift time
scales. Section~\ref{op} includes other possibilities to increase the drift
time scales. A detailed discussion of the theoretical background is given in
the Appendix A.

\section{Radial drift of the dust}\label{rdotd}

We consider a T Tauri star with a stellar mass of 1 $M_{\odot}$, a stellar
radius of 2.5 $R_{\odot}$ and a surface temperature of 4000 K. We assume that
the disk around this pre-main sequence star has an inner and an outer radius
of 0.03 AU and 150 AU \citep{Rod06}, respectively. The mass of the disk
$M_{\mathrm{disk}}$ is a free parameter of our model. We assume that the disk
is passive and irradiated by the central star under an angle of 0.05 rad. At a
disk radius of 1 AU these values imply a midplane temperature of 200 K
assuming that the disk is isothermal in the vertical direction. Moreover, we
will consider a turbulent disk. The amount of turbulence in the disk is
described by the turbulent $\alpha$ parameter \citep{ShaSun73}, which is a
free parameter of our model.

\begin{figure}
\begin{center}
\includegraphics[scale=.6]{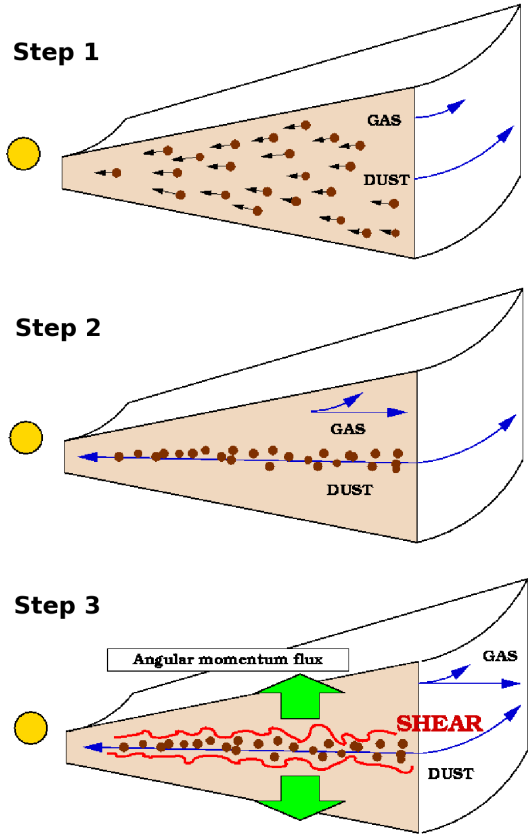}
\caption{This figure shows the three scenarios we will investigate in this
  paper. In step 1 we consider the drift of individual particles. In step 2 we
  investigate the influence of collective effects on the particle drift. These
  effects become of importance if the particles settle into a thin midplane
  layer. Finally, in step 3 we will investigate what happens if we include
  turbulent viscosity in our simulations. In this case, angular momentum is
  exchanged between the dusty midplane layer and the gaseous layers above the
  midplane. The horizontal arrows indicate the radial velocity of gas and
  dust.  The curved arrows indicate the azimuthal velocities.
\label{sketch}}
\end{center}
\end{figure}

Now, we will calculate the radial drift of the dust and the radial drift time
scales. In order to demonstrate the physics behind the calculations and its
implications on the radial velocities we will proceed in certain steps. In
every step more effects are included to demonstrate the influence on the drift
velocities (cf. Fig.~\ref{sketch}).
\begin{enumerate}
\item The first subsection addresses the drift of individual particles in a
gaseous disk. These results are valid when the dust and the gas are well
mixed.
\item However, under certain conditions (low-turbulence disks, large dust
particles), the dust sediments so close to the midplane that the density of
the dust exceeds the density of the gas. In that case collective effects of
the dust come into play, which is the topic of the second subsection.
\item Finally, in the last subsection, we also include the
effect of vertical angular momentum exchange between the dust midplane layer
and the gaseous layers above the midplane.
\end{enumerate}
We are primarily interested in drift time scales needed to keep mm to cm size
particles in the outer parts of the disk, hence, we will calculate
characteristic drift times at the very end of every of these three
subsections. We will see how these time scales are affected by including more
and more effects.

\subsection{Step 1 - Radial drift of individual particles}\label{subsec-singlepart}

\subsubsection{Equations}

The fundamental cause for inward drift of the dust is the difference in
velocity between gas and dust. While the dust moves with Keplerian velocity
the gas moves slightly sub-Keplerian. This is due to the fact that the gas is
not only affected by the gravitational and the centrifugal force but
additionally feels a radial pressure force that does not act on dust
particles. This extra force is caused by the decrease of gas pressure in the
radial direction.  Since this force, which exclusively acts on the gas, partly
compensates gravitation, the gas moves slower than Kepler speed and therefore
slower than any dust particle in the disk. Hence, the dust particle feels a
continuous headwind from the gas. This headwind causes the dust particle to
lose its angular momentum and spiral inward.

\cite{Whipple72} formulated the first equations for the radial drift of very
small and very large particles. \citet{Wei77} later derived a set of equations
with a general drag force to calculate the radial drift of solid particles of
any size. We will formulate all equations in the dimensionless Stokes number
formulation. The Stokes number can be regarded as a measure of grain size (see
Fig.~\ref{sta}). The full definition is given in Appendix \ref{thestokes}.  In
terms of this dimensionless formulation, these equations aquire the
form\footnote{The physical meaning of all variables is listed at the very end
of the paper.}
\begin{figure}
\begin{center}
\includegraphics[scale=.45]{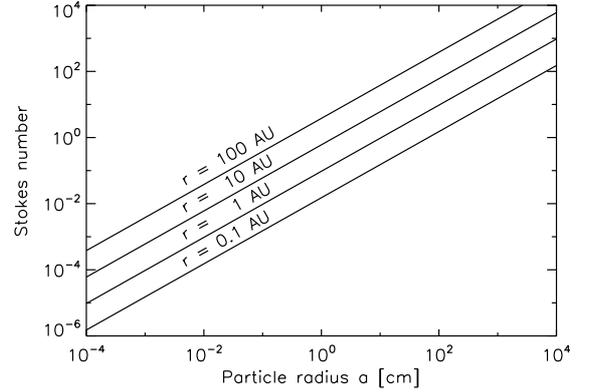}
\caption{The Stokes number St as a function of particle radius $a$ at
  different radii in the disk. In this calculation the solid particle density
  of the dust is 1.6 g/cm$^3$ and the stellar mass is 1 $M_{\odot}$. The
  innner and the outer radius of the disk are 0.03 AU and 150 AU,
  respectively. The disk mass is $10^{-2}M_{\odot}$.
\label{sta}}
\end{center}
\end{figure}
\begin{eqnarray}
0&=&\frac{w_{\varphi}}{\mbox{St}}+\frac{w_{r}}{2}\;,\nonumber\\
0&=&w_{\varphi}^2+v_{\mathrm{N}}w_{\varphi}+\frac{w_{r}^2}{4}\;.
\label{eq-weiden}
\end{eqnarray}
The variables $w_r$ and $w_{\varphi}$ denote the radial and azimuthal velocity
of the dust, respectively. The form of the drag law is implicitly included in
the Stokes number. The quantity $v_{\mathrm{N}}$ is the velocity by which the
gas moves azimuthally slower than Keplerian velocity $V_{\mathrm{k}}$,
i.e. $v_{\mathrm{gas}}=V_{\mathrm{k}}-v_{\mathrm{N}}$. The velocity
$v_{\mathrm{N}}$ will also turn out to be the maximum radial drift velocity of
the dust. We will take $v_{\mathrm{N}}$ from here on as our ''standard
velocity'' scale (cf. Eq.~\ref{etavkk}) apart from the Keplerian velocity
$V_{\mathrm{k}}$. For our disk model the quantity $v_{\mathrm{N}}$ is given by
\citep{Wei77,NakSekHay86}
\begin{equation}
v_{\mathrm{N}}=1.28\frac{c_{\mathrm{s}}^2}{V_{\mathrm{k}}}\;.
\end{equation}
In this expression $c_{\mathrm{s}}$ denotes the isothermal sound speed. The
number 1.28 is due to the specifications of our disk model
(cf. App. \ref{dismmodel}).  Note that the velocity scale $v_{\mathrm{N}}$
does not depend on the disk mass. In our disk model this quantity is also not
dependent on the location $r$ in the disk.

\begin{figure}
\begin{center}
\includegraphics[scale=.45]{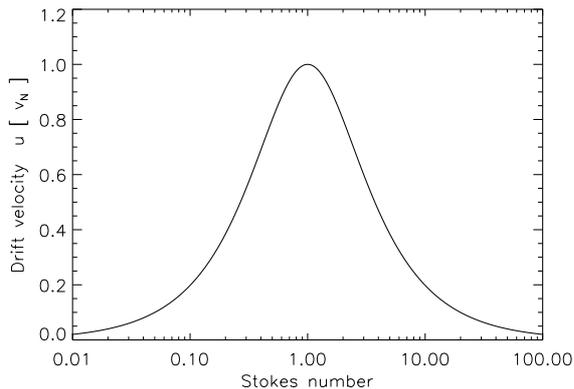}
\caption{The radial drift velocity of individual particles $w_r$ in units of
$v_{\mathrm{N}}$ as a function of Stokes number.\label{ipdpic}}
\end{center}
\end{figure}

The set of equations (\ref{eq-weiden}) is generally difficult to solve and
only numerical methods provide information about the drift velocity. However,
in some cases the situation simplifies. If the Stokes number is not dependent
on the particle velocity, the equations (\ref{eq-weiden}) can be solved
analytically. Assuming this independency a straightforward calculation yields
\begin{equation}
\label{vipd}
w_r=\frac{2}{\mathrm{St}+\frac{1}{\mathrm{St}}}v_{\mathrm{N}}\;.
\end{equation}
This equation directly shows that the drift velocity has a maximum when the
Stokes number is unity and the maximal drift velocity is $v_{\mathrm{N}}$. The
drift velocity as a function of Stokes number is shown in Fig.~(\ref{ipdpic}).

\subsubsection{Drift time scales for individual particles}

We will now investigate the radial drift times of individual particles in the
outer parts of the disk. More specifically we are interested in the conditions
on particle radius and particle porosity that provide time scales larger than
a few Myrs.

The drift timescale $\tdrift$ equals
\begin{equation}\label{eq-drift-onepart}
\tdrift=\frac{r}{w_r}\;,
\end{equation}
which should correspond to the age $\tage$ of the disks observed, thus a few
Myrs. Throughout the paper we focus on the radius of $r=100$ AU. To find the
critical Stokes numbers for which the radial drift time scale equals the age
of the disk, we replace $\tdrift$ in Eq.~(\ref{eq-drift-onepart}) by
$\tage$. We then insert Eq.~(\ref{vipd}) into Eq.~(\ref{eq-drift-onepart}),
and solve for St. This yields two critical Stokes numbers:
\begin{equation}
\mbox{St}_{\pm}=\frac{\tage v_{\mathrm{N}}}
{r}\pm \sqrt{\left(\frac{\tage v_{\mathrm{N}}}{r}\right)^2-1}\;.
\end{equation}
The interpretation of these two numbers is the following: If the Stokes number
of the dust particle falls into the interval $[\mbox{St}_{-},\mbox{St}_{+}]$,
then the drift timescale is shorter than $\tage$. If it falls outside of this
interval, then the drift time scale is long enough that these particles can be
observed in the protoplanetary disk of age $\tage$. Since the Stokes number
interval is a rather abstract depiction we reformulated it into a particle
radius interval with a similar meaning.

\begin{figure}
\begin{center}
\includegraphics[scale=.45]{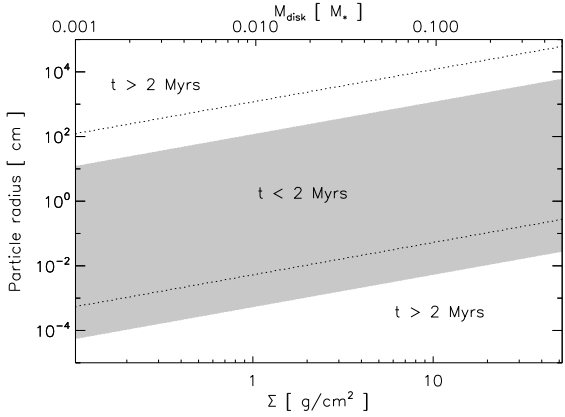}
\caption{The particle radius interval in which the individual drift time
scale at $r=$100 AU becomes shorter than 2 Myrs as a function of surface
density or equivalently disk mass (shaded region). The disk mass is computed
from the surface density using Eq.~(\ref{eq-diskmass}).  To illustrate the
effect of noncompact growth we also calculated the interval for a filling
factor of $10^{-1}$ (dotted lines).\label{ipdsurv}}
\end{center}
\end{figure}

The region of too short time scales at 100 AU is shown in Fig.~\ref{ipdsurv}
as a function of disk mass $M_{\mathrm{disk}}$ and surface density
$\Sigma$. In this diagram we applied a dust material density\footnote{10\%
silicate, 30\% carbonaceous material and 60\% ice} $\rho_{\mathrm{s}}=1.6$
g/cm$^3$, a Kepler frequency $\Omega_{\mathrm{ k}}=10^{-10}$/s and
$c_{\mathrm{s}}=2.6\times 10^{4}$ cm/s (corresponding to a temperature of 20
K). We take as the age of the disk $\tage=2$ Myrs. The maximum radial drift
velocity at this location is $v_{\mathrm{N}}=60$ m/s. We will make use of
these values at all times in this paper unless otherwise noted. The two Stokes
numbers $\mbox{St}_{\pm}$ that are implied by these values are
$\mbox{St}_{-}=0.002$ and $\mbox{St}_{+}=474$, corresponding to the lower and
upper edge of the grey zone in Fig.~\ref{ipdsurv} respectively.

The figure shows that the particle radius interval in which the time scale of
individual particles is shorter than 2 Myrs ranges over more than 5 orders of
magnitude in radius. Particles ranging from mm to cm in size are completely
included in this region independent of disk mass. The drift time scale of
sub-millimeter particles may exceed $\tage$ when the disk mass is higher than
$0.2$ $M_{\star}$.

The Stokes number as the crucial value for radial drift is not only affected
by particle radius but also by particle properties like porosity or fractal
growth \citep{Kem00}. This effect of noncompact growth may be considered by
introducing the filling factor of the particle $f$ defined by
$m_{\mathrm{p}}=V_{\mathrm{p}}\rho_{\mathrm{ s}}f$, where $m_{\mathrm{p}}$ and
$V_{\mathrm{p}}$ are the mass and the volume of the particle, respectively.
In Fig. (\ref{ipdsurv}) we also calculated the critical particle radius
interval for a filling factor of $f=10^{-1}$ (dotted lines).

The lower filling factor shifts the critical particle radius interval towards
higher particle radii. The drift time scale of mm size particles exceeds 2
Myrs when disk masses higher then 0.2 $M_{\star}$ are considered. For cm size
particles the time scale never exceeds 2 Myrs. For filling factors lower than
$10^{-3}$ the drift time scales of mm and cm size particles exceed 2 Myrs for
any disk mass higher than $10^{-3}$ $M_{\star}$. However, particles of mass 1
g and filling factors of $10^{-3}$ would imply a particle diameter of 5
cm. Since this particle size falls into the regime where compaction is thought
to occur \citep{BluWur99} this filling factor represents an unlikely case.
    
\subsection{Step 2 - Collective effects}\label{nakag}

The scope of the previous subsection can be expanded by including the
back-reactions from the dust to the gas. We do no longer consider a single
particle, but include how the entire swarm of dust particles can affect the
gas motion. The modified gas motion has the effect of reducing the rate by
which the gas extracts angular momentum from the dust, and thereby reduces the
radial drift of the dust. Such collective effects play the strongest role for
low $\alpha$-values so that a thin midplane dust layer can form in which the
dust density is high. This scenario of reduction of radial drift was described
by \citet{NakSekHay86}.

The necessity to take this additional effect into account may be illustrated
by regarding the following extreme scenario. We consider a hypothetical disk
in which the dust density is much higher than the gas density. In such a
dust-dominated disk the dust is hardly influenced by the gas. The gas, which
tends to move sub-Keplerian, is dragged along with the dust since it feels a
continuous tailwind of dust particles. Therefore, the gas perpetually gains
angular momentum from the dust and spirals outward.  The radial drift of the
dust is negligible because the dust-to-gas ratio is much higher than 1. This
is the reverse situation of the case described in Subsection
\ref{subsec-singlepart}. In general, though, we neither have a perfectly
dust-dominated nor gas-dominated situation. We then have to solve the
following set of equations \citep{NakSekHay86}
\begin{eqnarray}\label{eqnaka}
0&=&\;\;\; 2\Omega_{\mathrm{k}}u_{\varphi}-A\rho_{\mathrm{
d}}(u_r-w_r)\;-\frac{1}{\rho_{\mathrm{g}}}\partial_rp_{\mathrm{ g}}\nonumber\\
0&=&-\frac{1}{2}\Omega_{\mathrm{k}}u_r-A\rho_{\mathrm{ d}}
(u_{\varphi}-w_{\varphi})\nonumber\\ 0&=&\;\;\; 2\Omega_{\mathrm{
k}}w_{\varphi}-A\rho_{\mathrm{g}}(w_r-u_r)\;\nonumber\\
0&=&-\frac{1}{2}\Omega_{\mathrm{k}}w_r-A\rho_{\mathrm{
g}}(w_{\varphi}-u_{\varphi})\;.
\end{eqnarray}
The quantities $u$ and $w$ denote the velocity of the gas and the dust in a
Keplerian comoving frame, respectively. The subscripts $r$ and $\varphi$
indicate the radial and the azimuthal components of the velocities. The
variables $\rho_{\mathrm{g,d}}$ denote the mass densities of gas and dust and
the quantity $A$ is defined as
$A=\Omega_{\mathrm{k}}/\rho_{\mathrm{g}}\mathrm{St}$. We are primarily
interested in the radial dust velocity of the Nakagawa-Sekiya-Hayashi solution
(NSHs) which can be expressed as
\begin{equation}
\label{vcoll}
u^{\mathrm{NSH}}=\frac{2}{\psi\mathrm{St}+\frac{1}{\psi\mathrm{St}}} \psi
v_{\mathrm{N}}\qquad\mbox{and}\qquad\psi=\frac{1}{1+\epsilon}\;.
\end{equation}
The quantity $\epsilon\equiv \rho_d/\rho_g$ denotes the local dust-to-gas
ratio. When $\psi\rightarrow 1$, i.e., when the dust-to-gas ratio is
approaching zero, Eq.~(\ref{vcoll}) reduces to the corresponding equation for
single-particle drift, Eq.~(\ref{vipd}). The difference between these two
equations is the additional factor $\psi$ that modifies the Stokes number St
and the maximum drift velocity $v_{\mathrm{N}}$. Since $\psi$ is always
smaller than one, the collective radial drift of the dust will always be
smaller than the individual particle drift.

Taking collective effects into account requires knowledge about the dust
density. Therefore, certain disk parameters, i.e.\ the turbulence parameter
$\alpha$ as well as the initial dust-to-gas ratio $\epsilon_0$ before
sedimentation, become of importance. Another dimensionless number, which now
comes into play, is the turbulence parameter $q$. This number determines
whether turbulent diffusion is realised by small turbulent eddies moving fast
or by big eddies moving slow (cf. Appendix \ref{apres}). These quantities
determine the thickness of the dust layer $h$ and, hence, the dust density
(cf. Eqs.~\ref{gp} and \ref{rhodust}). In these equations we assume that the
dust density in the vertical direction has a gaussian shape. This ansatz might
be put into question if the turbulence is self-induced \citep{Wei79}. Although
\cite{jhk06} showed that the vertical dust density in self-induced turbulence
has a gaussian shape for canonical dust-to-gas ratios, the vertical structure
can show a different shape especially when $\epsilon$ is increased for
instance through photoevaporation \citep{Wei06}.

\begin{figure}
\begin{center}
\includegraphics[scale=0.12]{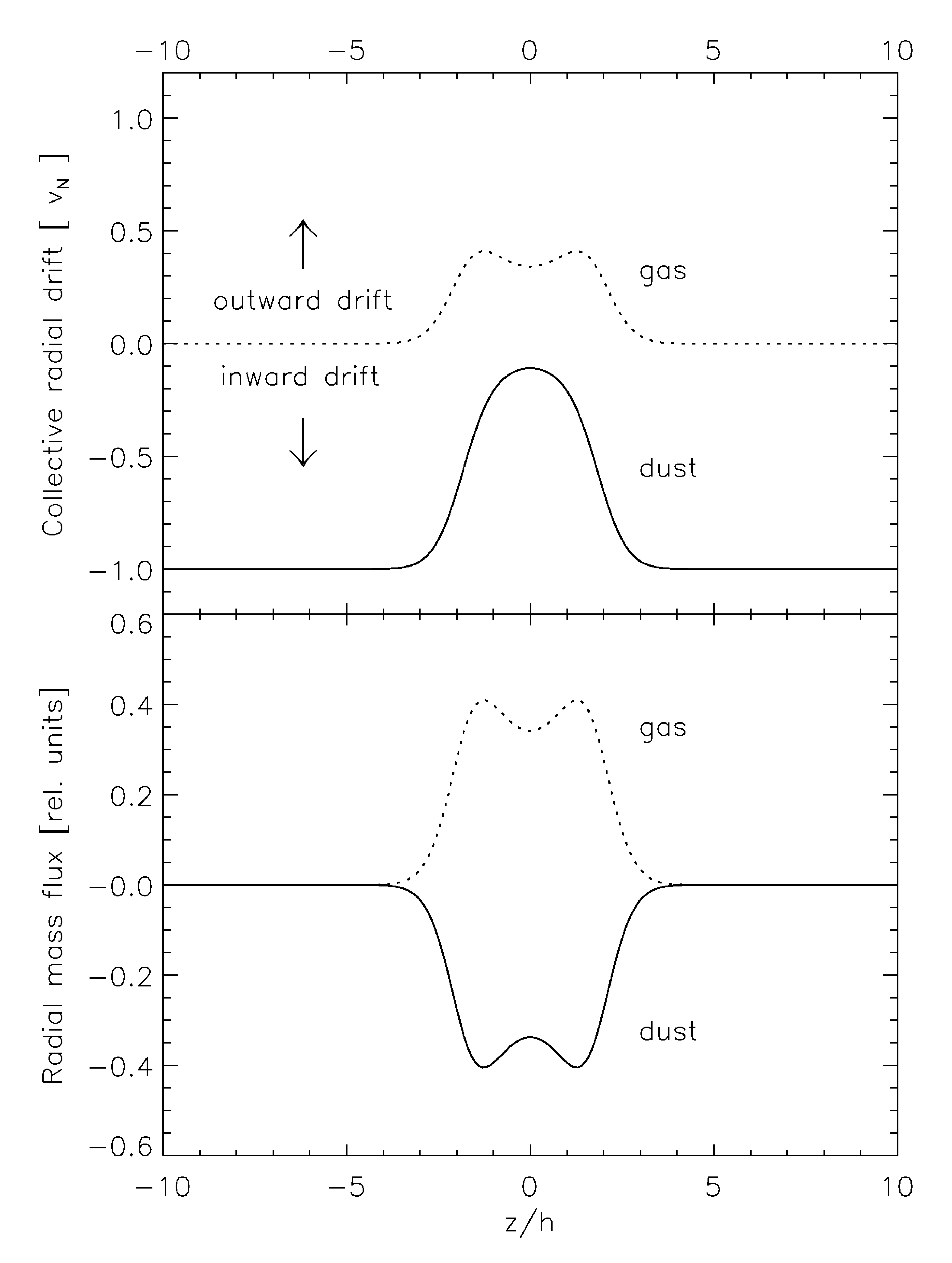}
\caption{The upper figure shows the collective radial velocities of gas and
  dust of the laminar NSH solution in terms of $v_{\mathrm{N}}$ as a function
  of height above the midplane. The lower figure shows the radial mass flux of
  gas and dust in arbitrary units. The values applied in this calculation are
  $\mbox{St}=1$, $\alpha=10^{-5}$, $q=1/2$ and
  $\epsilon_0=10^{-2}$.\label{naka}}
\end{center}
\end{figure}

Since the dust density is a function of height above the midplane $z$, the
radial drift velocities are dependent on $z$ as well. This vertical dependency
is shown in Fig.~(\ref{naka}) for an exemplary NSH solution. In this
calculation we applied the values $\mbox{St}=1$, $\alpha=10^{-5}$, $q=1/2$ and
$\epsilon_0=10^{-2}$.

The plot shows that the dust moves inwards while the gas moves outwards which
is generally the case in the NSH solution. In the higher regions of the disk
$(\,|z|>3\;h\,)$ the dust-to-gas ratio is much smaller than unity causing the
collective drift behaviour to match the individual particle drift. However,
closer to the midplane of the disk collective effects become important. With
increasing dust-to-gas ratio towards the midplane, the radial inward drift of
the dust decreases while the gas starts to move outwards. The clear difference
in velocities between the collective drift and the individual drift around the
midplane along with the fact that most of the dust is located in this region
demonstrates the importance of collective effects for disks with low
turbulence.

\begin{figure}
\begin{center}
\includegraphics[scale=.45]{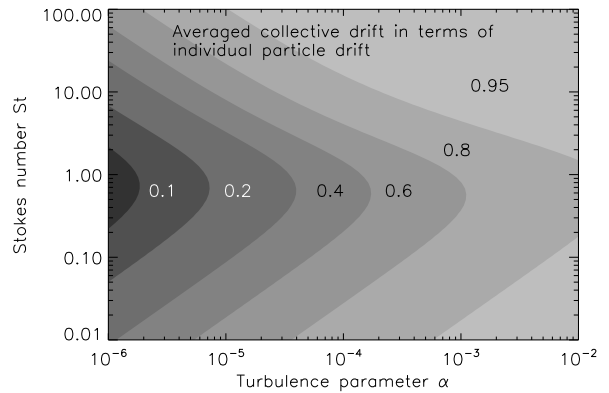}
\caption{A contour plot of the vertically averaged collective radial drift
velocities of the dust in terms of the individual particle drift velocity as a
function of turbulence parameter $\alpha$ and Stokes number St. The numbers in
the diagram indicate the contour level and are related to the line on the
left, respectively. The parameters for this calculation are $q=1/2$ and
$\epsilon_0=10^{-2}$.\label{naka2}}
\end{center}
\end{figure}

The radial velocities as a function of height above the midplane do not
directly tell something about the entire radial flow of the dust since the
dust itself is vertically distributed in a certain way. For this reason we
will now calculate the vertically averaged radial velocity of the dust. This
integrated velocity is given by
\begin{equation}
\label{netv}
\bar{u}^{\mathrm{NSH}}=\frac{1}{\Sigma_{\mathrm{d}}}
\int_z\rho_{\mathrm{d}}(z)u^{\mathrm{NSH}}(z)\;\mbox{d}z\;.
\end{equation}
A contour plot of this quantity as a function of the turbulence parameter
$\alpha$ and the Stokes number St is shown in Fig. (\ref{naka2}).  The drift
velocities in this diagram were expressed in terms of the corresponding
individual particle drift velocity in order to explicitly point out the
differences between these two models.

The figure shows that for fixed Stokes numbers the deviation increases
continuously with lower turbulence in the disk. Lower $\alpha$ values imply
thinner dust layers and, therefore, higher dust-to-gas ratios. With higher
dust-to-gas ratios the back reaction of the dust to the gas increases, and
hence the deviation between individual and collective drift velocities.  One
obvious solution to the whole radial drift problem of grains in the outer
parts of the disk would be to continuously decrease the amount of turbulence
in the disk or even to set $\alpha$ to zero. However, \cite{Wei79} has shown
that a shear-instability between the dust layer and the gas induces a weak,
but non-negligible level of turbulence. This turbulence is called
`self-induced turbulence' which constrains the $\alpha$-value to be at least
of the order of $\sim 10^{-6}$ (cf. Appendix \ref{sia}).

For fixed $\alpha$ Fig.~\ref{naka2} shows that for low Stokes numbers (small
grains) the drift behavior approaches the individual particle drift. Low
Stokes numbers imply thick dust layers, causing low dust-to-gas ratios. For
high Stokes numbers (large grains), very thin dust layers are obtained. One
would intuitively think that this maximizes the collective effects. However,
as can be seen from Eq.~(\ref{vcoll}), in the limit of St $\rightarrow\infty$
one gets $u^{\mathrm{NSH}}\rightarrow 2 v_{\mathrm{N}}/\mbox{St}$ which is
equal to the individual particle drift of Eq.~(\ref{vipd}). So for large St
the radial drift indeed drops, but not due to collective effects.

For a Stokes number of unity and a turbulence parameter of $10^{-6}$ the
dust-to-gas ratio in the midplane is given by
\begin{equation}
\epsilon_{\mathrm{mid}}=
\epsilon_0\frac{H}{h}=\epsilon_0\sqrt{\frac{\mbox{St}}{\alpha}}=10.
\end{equation}
Therefore, the collective radial drift in the midplane in terms of the
individual particle drift according to Eq.~(\ref{vcoll}) is
$u^{\mathrm{NSH}}_{\mathrm{mid}}/v_{\mathrm{ N}}=(1+\epsilon)^{-2}=0.008$.
However, the vertically averaged drift velocity of the dust in terms of the
individual particle drift in Fig.~(\ref{naka2}) is only 0.07, which is almost
one order of magnitude higher. The reason for this is that the largest radial
dust mass flux is not in the midplane, but slightly above the midplane (see
Fig.~\ref{naka}). The mass flux is the product of dust density
$\rho_{\mathrm{d}}$ and dust radial velocity $u^{\mathrm{NSH}}$. Although
$\rho_{\mathrm{d}}$ drops strongly slightly above the midplane, the radial
velocity $u^{\mathrm{NSH}}$ increases even faster, so that the product
$\rho_{\mathrm{ d}}\;u^{\mathrm{NSH}}$ has a maximum slightly above the
midplane. While the formation of a dense dust layer can reduce the radial
drift velocity in the midplane by a factor of 100, the vertically averaged
radial drift velocity can be only reduced by a factor of at most 10.

\subsubsection{Radial drift times including collective effects}

Armed with the above drift velocity expressions we now calculate the
conditions on particle radius and particle porosity that provide time scales
larger than 2 Myrs taking into account collective effects. 
\begin{figure}
\begin{center}
\includegraphics[scale=.45]{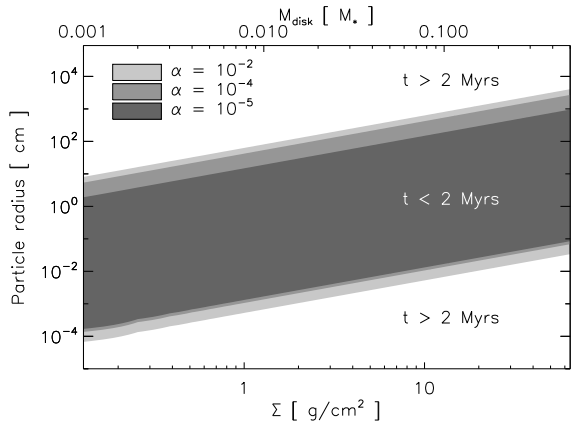}
\caption{As Fig.~\ref{ipdsurv}, but now with collective effects of the dust
and the gas included, at $r=100$ AU. Different grey scales are the results for
different levels of turbulence, hence different thicknesses of the dust
midplane layer. Note that for $\alpha=10^{-2}$ the results are virtually
identical to the single-particle case (Fig.~\ref{ipdsurv}), since in this case
the dust layer is so thick that the dust-to-gas ratio is much less than unity.
\label{coll1}}
\end{center}
\end{figure}

At first we focus on the conditions on the particle radius. The interval of
this particle property that corresponds to time scales shorter than 2 Myrs is
shown in Fig.~(\ref{coll1}) for different $\alpha$-parameters. The second
turbulence parameter $q$ is fixed at $1/2$ at all times, the initial
dust-to-gas ratio is $10^{-2}$ and the filling factor $f$ is unity. All other
parameters were already mentioned in the last section and are not changed
throughout the paper unless directly stated.

According to this plot, the critical particle radii that provide the requested
time scales hardly differ from the critical particle radii of the individual
particle drift calculated in the last section. Even for small turbulence
parameters which favour collective effects the time scales for mm to cm size
particles are shorter than 2 Myrs for any disk mass considered. The reason for
this is that for very high and very low Stokes numbers, like the two critical
numbers St$_{-}$ and St$_{+}$ representing the boundaries of the grey areas in
Fig~\ref{coll1}, collective effects play a minor role (see Fig.~\ref{naka2},
and discussion in the last subsection). The Stokes numbers for which the
collective effects play the strongest role lie in the middle of these grey
areas, i.e.~where the drift time scales are anyway much too short to be
compatible with the observations of mm-sized particles in protoplanetary
disks.

\begin{figure}
\begin{center}
\includegraphics[scale=.45]{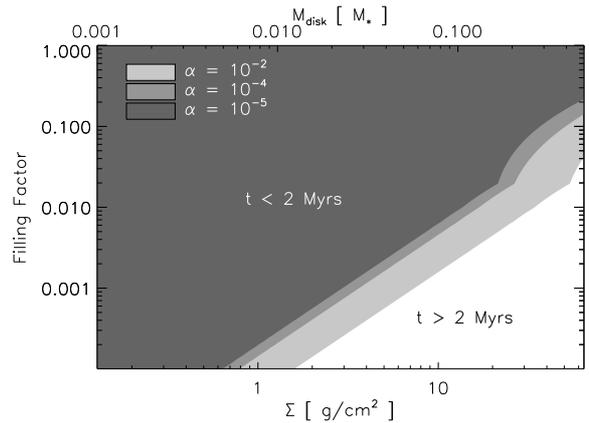}
\caption{The dust particle filling factor that provides time scales larger
than 2 Myrs as a function of disk mass for different turbulence parameters at
$r=100$ AU in the disk. In this calculation collective effects of dust and gas
are taken into account.\label{coll2}}
\end{center}
\end{figure}

So what about the effects of fractal or porous growth? For simplicity we set
the mass of the dust particle to be 1 g and then calculate the particle
filling factor that provides time scales larger than 2 Myrs. For a filling
factor of unity a dust particle of 1 g corresponds to a particle radius of 1/2
cm. For $f=10^{-4}$ the particle radius can be calculated to be 11 cm. Like in
the last paragraph we will perform the time scale calculation in dependency on
the disk mass. The results of this calculation is shown in Fig.~(\ref{coll2})
for different values of the turbulence parameter $\alpha$.
This diagram shows that for filling factors lower than $10^{-2}$ the time
scale exceeds 2 Myrs subject to the condition that the disk mass is higher
than $\sim 0.2$ $M_{\star}$. For even higher disk masses the filling factor
may exceed 0.1 for certain turbulent $\alpha$ parameters. This filling factor
corresponds to a particle radius of 1 cm. \cite{ormel07} showed that particle
growth in protostellar disks can be associated with filling factors of less
than $\sim 10^{-1}$. Therefore fractal growth seems to be an actual
possibility to considerably increase the radial drift time scales of the
dust. We will come back to that point in the discussion.

\subsection{Step 3 - Effect of turbulent viscosity}

We will now investigate the role of turbulent viscosity in addition to the
effects studied so far. Including viscosity terms will have the opposite
effect on the drift velocities than the collective effects. It will increase
the radial drift of the dust and shorten the drift time scales.  We will give
the Navier-Stokes equation (NSe) including collective effects and viscosity
terms and solve these equations numerically. However, we would like to discuss
first why turbulent vicosity increases the radial drift of the dust.  Under
certain conditions, i.e.\ small turbulence parameters $\alpha$ or high Stokes
numbers St, the previous sections have shown that the dust may settle into a
thin midplane layer. When the dust-to-gas ratio inside this layer exceeds
unity the gas is dragged along with the dust. Both components, dust and gas,
tend to move with Keplerian velocity.

On the other hand, above the dust layer the dust-to-gas ratio is much smaller
than unity. In this region the dust particles feel a continuous head wind
which forces them to move with slightly sub-Keplerian velocity. This vertical
decrease in azimuthal velocity from Keplerian velocity in the midplane to
sub-Keplerian velocity in the higher regions of the disk produces a nonlinear
velocity gradient in both the gas and the dust.

If we include viscosity in our considerations it attempts to damp nonlinear
spatial velocity differences. The vertical velocity gradient described above
represents such a difference. Turbulent viscosity now acts in such a manner
that it transports angular momentum from the midplane to the higher regions of
the disk. While the midplane, the region where most of the dust is located,
loses angular momentum and falls inward, the regions above the midplane gain
angular momentum and move outward (cf. Fig.~\ref{sketch}). This mechanism of
vertical angular momentum exchange was first investigated by \cite{YouChi04}.
 
\subsubsection{Navier-Stokes equations}

The Navier-Stokes equations for this problem are basically the set of
equations (\ref{eqnaka}) plus some second order derivative terms due to the
inclusion of viscosity
\begin{eqnarray}
\label{nstokes}
0&=&2\Omega_{\mathrm{k}}u_{\varphi}\quad-A\rho_{\mathrm{
d}}(u_r-w_r)\;+\;\nu_{\mathrm{
g}}\partial_z^2u_r-\frac{1}{\rho_{\mathrm{g}}}\partial_rp_{\mathrm{
g}}\nonumber\\ 0&=&-\frac{1}{2}\Omega_{\mathrm{k}}u_r-A\rho_{\mathrm{
d}} (u_{\varphi}-w_{\varphi})+\;\,\nu_{\mathrm{
g}}\partial_z^2u_{\varphi}\nonumber\\ 0&=&2\Omega_{\mathrm{
k}}w_{\varphi}\;\;\,-A\rho_{\mathrm{g}}(w_r-u_r)\;+\;\,
\frac{\nu_{\mathrm{d}}}{\rho_{\mathrm{
d}}}\partial_z\left(\rho_{\mathrm{d}}\partial_zw_r\right)\nonumber\\
0&=&-\frac{1}{2}\Omega_{\mathrm{k}}w_r-A\rho_{\mathrm{
g}}(w_{\varphi}-u_{\varphi})+\; \frac{\nu_{\mathrm{d}}}{\rho_{\mathrm{
d}}}\partial_z\left(\rho_{\mathrm{d}}\partial_zw_{\varphi}\right)\;.
\end{eqnarray}
Hence, the algebraic Eqs.~(\ref{eqnaka}) turn into four coupled, differential
equations of second order. The left hand side of the Navier-Stokes equations
representing the time dependencies are set to be zero since we are interested
in steady state solutions. The vectors $\vec{u}$ and $\vec{w}$ denote the
velocities of the gas and the dust, respectively. The first terms on the right
side correspond to the Coriolis force. These terms arise from the fact that
the equations are formulated in a comoving frame. The second term represents
the drag force coupling between the gas and the dust. The effects of viscosity
show up in the third terms. The expressions for the viscosity of the gas and
the dust can be found in Appendix \ref{tb}. In the following we will denote
viscosity terms which involve derivatives of radial (azimuthal) velocities as
'radial (azimuthal) viscosity terms'. The very last term in the first line is
an extra force acting on the gas which is caused by a radial pressure
gradient. This term is responsible for the gas moving slower than the dust and
causes the radial drift. The densities of gas and dust serve as input for the
Navier-Stokes equations.

\citet{TakLin02} also investigated the effect of gas viscosity on the drift of
dust particles, but they neglected collective effects. This allowed to solve
the equations analytically. The drift of the dust particles in their
calculations was a superposition of two different effects: The individual dust
particle velocity with respect to the gas and the velocity of the gas itself.

The former part of the dust particle drift was discussed in detail in Section
\ref{subsec-singlepart}. The second part of the dust particle drift
investigated by Takeuchi and Lin was due to the gas accretion process. This
process of the gas is associated with a certain radial accretion
velocity. Since the dust is to some extent coupled to the motions of the gas
the dust is carried along with the accreting gas which leads to an extra
source of radial particle drift.

In the beginning of this section we described that gas viscosity also
increases the radial drift of the dust when collective effects come into play,
i.e. when the dust settles into a thin midplane layer and starts to affect the
motions of the gas. This process is different from the single particle
considerations discussed by Takeuchi and Lin since it is caused by collective
effects and not by gas accretion. In the following, we will estimate the ratio
of these two radial drift velocities.

The additional drift due to the accretion process may be estimated by a
characteristic accretion velocity of the gas which is given by
$v_{\mathrm{acc}}\propto\alpha c_{\mathrm{ s}}^2/V_{\mathrm{k}}$
\citep{ShaSun73}. Viscous collective effects imply drift velocities of order
$v_{\mathrm{coll}}=c_{\mathrm{ s}}^3/V_{\mathrm
k}^2\epsilon_0\mbox{Re}^{\star}$ \citep{Wei03}. The ratio
\begin{equation}
\xi=\frac{v_{\mathrm{acc}}}{v_{\mathrm{coll}}}=\alpha\epsilon_0\mbox{Re}^{\star}\frac{V_{\mathrm{k}}}{c_{\mathrm{s}}}
\end{equation}
has values of at most $\approx 10^{-1}$ for\footnote{Estimated values at 100
AU} $\alpha=10^{-2}$, $\epsilon_0=10^{-2}$, $\mbox{Re}^{\star}=10^2$,
$V_{\mathrm{k}}=3\times 10^5$ cm/s and $c_{\mathrm{s}}=3\times 10^4$ cm/s. For
smaller turbulence parameters which we will consider in this paper the
influence of gas accretion will be even lower. 

\subsubsection{Numerical results}

\begin{figure}
\begin{center}
\includegraphics[scale=.18]{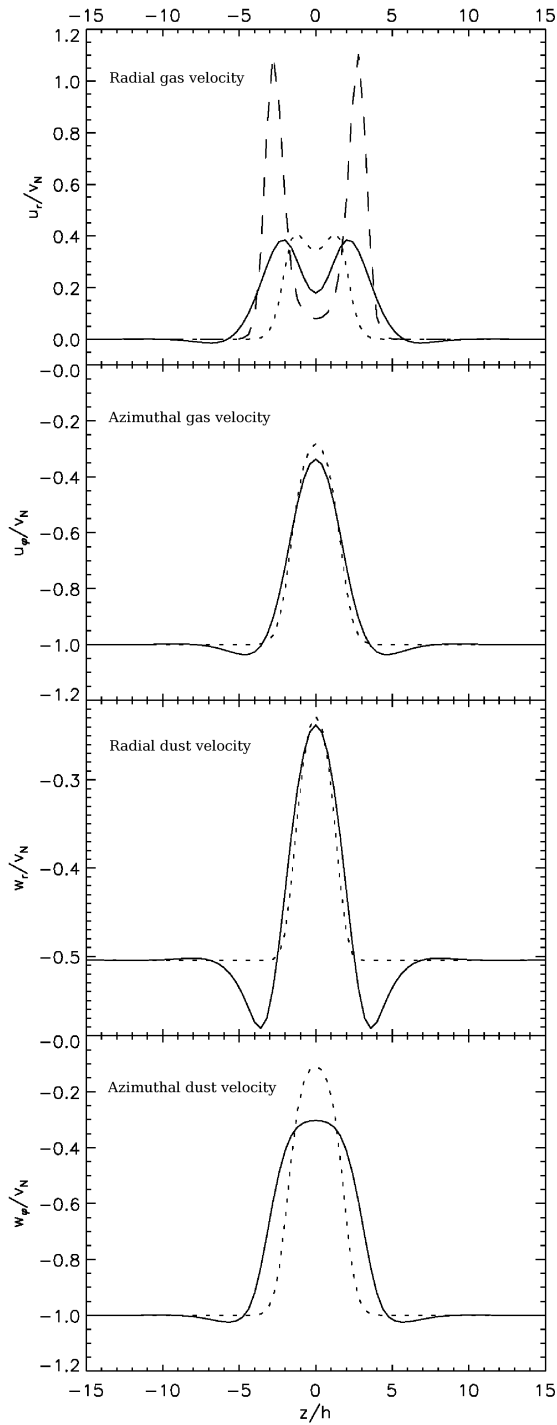}
\caption{The azimuthal and radial gas and dust velocities as a function of
  height above the midplane. The dotted lines denote the analytical normalised
  NSH solution without viscosity ($\nu=0$) as discussed in Section
  \ref{nakag}. The solid lines indicate the numerical solution of the
  Navier-Stokes equations including azimuthal and radial viscosity terms. The
  dashed line in the top diagram shows the radial velocity of the gas if
  radial viscosity terms are neglected. The values for this simulation are
  $\mbox{St}=1.2$, $\alpha=10^{-6}$, $q=0.5$ and $\epsilon_0=0.01$. Note that
  more than 80\% of the dust in within the $z-$interval $[-2h,2h]$.\label{s1}}
\end{center}
\end{figure}

The parameters for the simulation are $\mbox{St}=1.2$, $\alpha=10^{-6}$,
$q=1/2$, $\epsilon_0=10^{-2}$. The results are shown in Fig.~\ref{s1}. The two
parameters $\alpha$ and St are chosen in a way that effects of viscosity
become visible. These values imply a dust-to-gas ratio of 5 in the midplane of
the disk and a half thickness of the dusty midplane layer which is $\approx
0.002$ $H$. The turbulent motions of the gas have a speed of $0.005$
$v_{\mathrm{{N}}}$. The dotted lines in the Fig.~\ref{s1} indicate the
analytical solution of the laminar NSH equations ($\nu=0$) which were
discussed in the last section.  The solid lines in Fig.~\ref{s1} indicate the
numerical solution including viscosity which differs significantly from the
NSH solution.

Let us focus on the radial dust velocity since we are primarily interested in
radial drift time scales. The radial flow of the dust is significantly
affected by turbulent viscosity if $\alpha$ is smaller than $10^{-4}$. In this
regime the effect is largest for Stokes numbers $\approx 5$. For
$\alpha>10^{-4}$ the radial flow is approximately the flow predicted by
Nakagawa et al. and viscosity seems to play a minor role. The radial dust
velocities in the midplane with and without viscosity terms may differ by a
factor of 5 for small $\alpha$ parameters and $\mbox{St}\approx 5$.

The azimuthal dust and gas velocities as a function of height above the
midplane vary in a complex manner. However, for Stokes numbers smaller than
unity the situation with regard to the azimuthal velocities simplifies. In
this regime these velocities do hardly differ from the expression given by
Nakagawa et al. and viscosity seems to be negligible.

The radial outflow of the gas, which is shown in Fig.~(\ref{s1}), is reduced
if turbulent viscosity is included. This decrease may be up to a factor of 30
for small St and $\alpha$ parameters. For turbulence parameters higher than
$10^{-4}$ the radial net outflow of the gas differs less than 10\% from the
outflow predicted by the NSH equations.

\subsubsection{Width of the azimuthal gas velocity layer}\label{wii}

The calculation of radial drift velocities in protostellar disks including
collective effects and effects of viscosity are a challenging topic. Most
equations can not be solved analytically and only numerical solutions provide
information on the evolution of these disks. Therefore, disk model
simplifications often come into play.

One simplification is that the dust sub-disk is assumed to be extremely thin
and thought to behave, to some extent, like a solid disk. This approximation
is called ''plate drag approximation'' \citep{GolWar73}. Under this condition,
the gas layer above the dust layer can be described by an Ekman layer: The gas
in the midplane is forced to move along with the Keplerian rotating solid
equatorial subdisk. High above the midplane the gas is in equilibrium with the
radial gas pressure gradient, yielding a slightly sub-Keplerian rotational
velocity. The Ekman layer is the transitional region between these two
extremes. The thickness of this layer depends on the viscosity of the gas.

In this subsection, we will compare our results with the predictions of the
simplified model described above. We want to know the extent of the region
where gas and dust affect each other and effects of viscosity become of
importance. The comparison with regard to the drift velocities implied by this
approximation, however, will be discussed in Section \ref{platte}.

\begin{figure}
\begin{center}
\includegraphics[scale=.45]{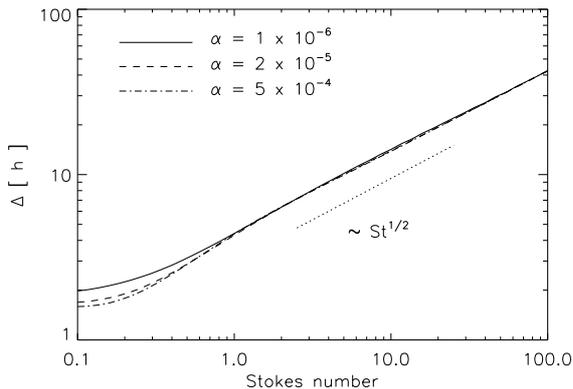}
\caption{The width of the azimuthal gas velocity distribution in units of the
dust scale height $h$ as a function of Stokes number St for 3 different
turbulence parameters $\alpha$. The dotted line indicates the
$\sim\mbox{St}^{1/2}$ dependency of the Ekman layer.\label{upa}}
\end{center}
\end{figure}

To quantify the length scale over which viscous collective effects play an
important role, we define a measure $\Delta$ by
\begin{equation}
\Delta(\alpha,\mbox{St})=\int g(z)|z|\;dz.    
\end{equation}
The function $g(z)$ is given by 
\begin{equation}
g(z)=c_{\mathrm{N}}|u_{\varphi}+v_{\mathrm{N}}|z^2\;.
\end{equation}
The constant $c_{\mathrm{N}}$ provides the normalization of $g$. With this
distribution function, deviations from the single particle solution
$u_{\varphi}+v_{\mathrm{N}}$ are weighted in a way that differences high above
the midplane are more important than differences close to the
midplane. Therefore, $\Delta$ provides informations about the width of the
vertical azimuthal velocity distribution of the gas. The dependence of this
quantity as a function of St is shown in Fig. (\ref{upa}) for 3 different
$\alpha$-values.

According to this diagram the value of the length scale $\Delta$ is a few dust
scale heights as long as the Stokes number is smaller than unity. For higher
St values this quantity increases exponentially up to more than 40 $h$ for
$\mbox{St}=100$. We also find that $\Delta$ is hardly dependent on the
turbulence parameter $\alpha$. This behaviour may be understood by
investigating the length scale of the classical Ekman layer
\begin{equation}
\frac{\Delta_{\mathrm{E}}}{h}\sim\frac{1}{h}\sqrt{\frac{\nu_{\mathrm{g}}}{\Omega_{\mathrm{k}}}}\sim\sqrt{\alpha}\frac{H}{h}\sim\sqrt{\mbox{St}}\;.
\end{equation}
The dotted line in Fig. (\ref{upa}) indicates this dependency which shows that
the gas layer in fact acts like an Ekman layer when the Stokes number exceeds
unity.

\subsubsection{Vertical flow of angular momentum}

A remarkable effect of the viscosity is the radial inward drift of the gas
which is impossible in the laminar NSH solution. At certain heights above
the midplane the gas moves inwards (see Fig. \ref{s1} at $z=\pm 8h$ for
example).

To understand this effect, we provide the basic scenario. The gas in the
midplane of the disk dragged by the dust moves azimuthally faster than the gas
outside the dust layer, which causes a vertical velocity gradient. Since
viscosity tries to equalize such velocity gradients the gas in the higher
regions of the disk is accelerated, decelerating the gas in the
midplane. Therefore, viscosity transports angular momentum from the midplane
to the higher regions of the disk.

\begin{figure}
\begin{center}
\includegraphics[scale=.45]{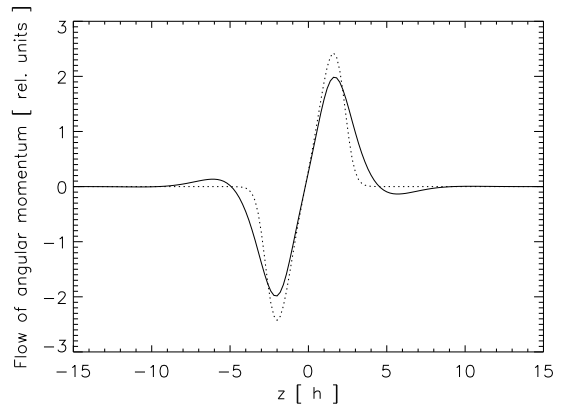}
\caption{This graph shows $\partial_zu_{\tiny\varphi}$ which indicates the
flow of angular momentum in the vertical direction. The solid and the dotted
lines show the flow of momentum with and without radial viscosity terms,
respectively. The azimuthal viscosity terms are included at all
times.\label{angmom}}
\end{center}
\end{figure}

To substantiate this effect we calculate the flow of angular momentum of the
gas in the vertical direction. The structure of this flow can be analysed by
calculating $\partial_zu_{\tiny\varphi}$ (see Fig.~\ref{angmom}). This
calculation was performed with the same parameter values as used in the last
section.

The results show that the maximum vertical upward flow of angular momentum
takes place at $\pm 2h$. It also shows that there is a vertical downward flow
of angular momentum at about $\pm 6h$. This flow is strongly associated with
the radial inward drift of the gas at certain heights of the disk and the
inclusion of radial viscosity terms. This behaviour can be understood by
performing simulations without radial viscosity terms:

If only azimuthal viscosity terms are included the azimuthal gas velocities
continuously decrease with increasing distance from the midplane. This means
that angular momentum is generally transported in the higher regions of the
disk and never towards the midplane. This suggests that the vertical downward
flow of angular momentum is an effect caused by radial viscosity. To
substantiate this assumption Fig. (\ref{angmom}) also shows the vertical flow
of angular momentum when radial viscosity terms are neglected (dotted
lines). The inflow vanishes in this case.

We also calculated the radial velocity of the gas without radial viscosity
terms included. The results of this calculation indeed demonstrate that the
radial inward drift of the gas vanishes in this case (see Fig. \ref{s1}). The
results also show the occurrence of two narrow peaks in the vertical velocity
distribution of the radial gas velocity without radial viscosity terms. These
peaks imply high velocity gradients. Radial viscosity, once included in the
simulation, reduces these velocity differences by radially accelerating the
neighbouring regions. This acceleration leads to a decrease in the azimuthal
velocities since $\dot{u}_{\varphi}\sim-u_{\mathrm{r}}$ due to Coriolis
forces. This again causes the gas to drift inward.

Figure \ref{s1} also shows that the radial outflow of the gas may be faster
than $v_{\mathrm{N}}$ if radial viscosity terms are neglected. The azimuthal
velocity differences in gas and dust that initially cause any drift behaviour
are of the same order of magnitude. Therefore, it appears unjustified to
neglect radial viscosity terms as often implicitly done by using the plate
drag approximation for example. 

\subsubsection{Integrated radial velocities}\label{irvv}

\begin{figure}
\begin{center}
\includegraphics[scale=.45]{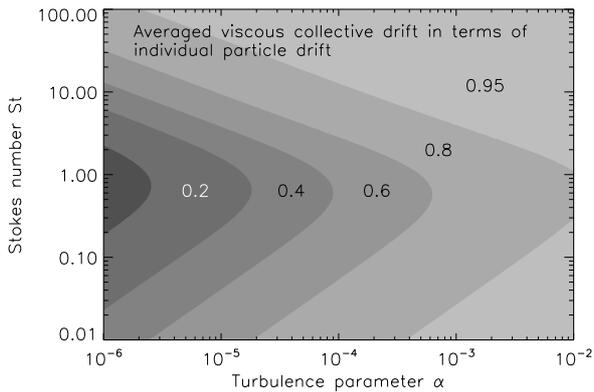}
\caption{A contour plot of the integrated radial dust velocity
$\bar{u}^{\mathrm{V}}$ under the influence of viscosity as a function of
turbulence parameter $\alpha$ and Stokes number St. The numbers related to the
contour lines on the left side, respectively, indicate the net drift velocity
in units the individual particle drift $u$.\label{result}}
\end{center}
\end{figure}

We now calculate the net flow of the dust $\bar{u}^{\mathrm{V}}$ according to
Eq.~(\ref{netv}). The result is shown in Fig. (\ref{result}) expressed in
terms of the individual particle drift velocity.
According to these results the drift behaviour for high turbulence parameters
is that of individual particles and neither collective effects nor effects of
viscosity seem to play a major role in this part of the diagram. The net dust
velocity has values of about $-v_{\mathrm{N}}$ for $\mbox{St}\approx 1/2$
(cf. Fig.~\ref{result}) and decreases with lower $\alpha$ values and with
growing distance from $\mbox{St}\approx 1/2$.

\begin{figure}
\begin{center}
\includegraphics[scale=.45]{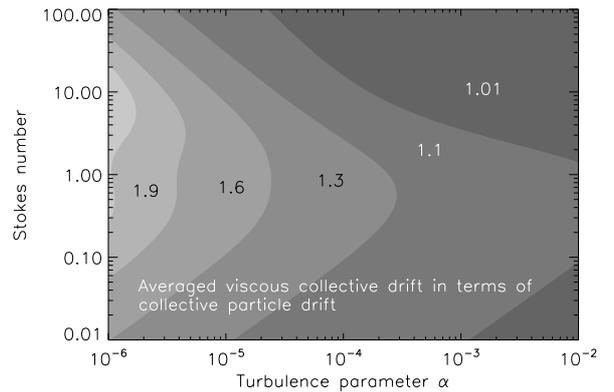}
\caption{A contour plot of the integrated radial dust velocity
$\bar{u}^{\mathrm{V}}$ under the influence of viscosity as a function of
turbulence parameter $\alpha$ and Stokes number St. The numbers related to the
contour lines on the left side, respectively, indicate the net drift velocity
in units of the NSH drift $\bar{u}^{\mathrm{NSH}}$ without
viscosity.\label{res2}}
\end{center}
\end{figure}

To demonstrate how viscosity changes the collective drift behaviour
investigated in step 2, it is suggestive to express the viscous collective
drift $\bar{u}^{\mathrm{V}}$ in terms of the NSH drift
$\bar{u}^{\mathrm{NSH}}$. A contour plot of this ratio can be seen in
Fig. (\ref{res2}). This plot shows that the radial velocities calculated in
this section exceed the drift due to collective effects by a factor of 2 at
most if very low turbulence parameters are considered. For $\alpha$ parameters
higher than $10^{-4}$ viscosity alters the drift scales by a factor of 1.2 in
the most extreme case. The deviation from individual particle velocities due
to collective effects were more pronounced than those due to
viscosity. Therefore, we conclude that the drift behaviour is predominantly
determined by collective effects and not by effects of viscosity.

\begin{figure}
\begin{center}
\includegraphics[scale=.45]{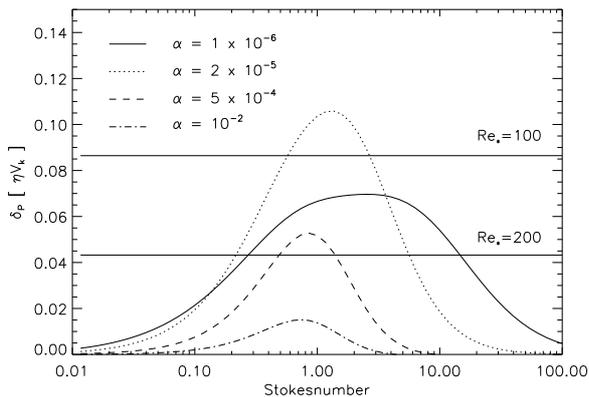}
\caption{The radial velocity deviation induced by the inclusion of viscosity
as a function of Stokes number St for different turbulence parameters
$\alpha$. The horizontal lines indicate the viscous radial drift predicted by
the plate drag approximation for two different critical Reynolds
numbers.\label{pd}}
\end{center}
\end{figure} 
 
\subsubsection{Plate drag approximation}\label{platte}

Here we will compare our results with previous work. We will consider the
predictions of the ''plate drag approximation''. We would like to investigate
if these two drift models predict the same radial velocities in certain
parameter regimes.

In the plate drag approximation the drift induced by viscosity is given by
\citep{GolWar73,Wei03}
\begin{equation}
u_{\mathrm{PD}}=\frac{\eta
V_{\mathrm{k}}}{c_{\mathrm{s}}\epsilon_0\mbox{Re}_{\star}\sqrt{2\pi}}\;.
\end{equation}
The derivation of this expression is based on the assumption that the dust
sublayer behaves like a solid disk subject to viscous stress on its surface by
a turbulent boundary layer. This stress extracts angular momentum from the
dust layer which implies radial drift of the dust \citep{Wei06}. The quantity
$\mbox{Re}_{\star}$ denotes the critical Reynolds number which indicates the
transition point between laminar and turbulent flow. The value of this number
depends on the geometry of the flow and is usually determined
experimentally. We calculated the viscous drift for two different critical
Reynolds numbers $\mbox{Re}_{\star}=100$ and $\mbox{Re}_{\star}=200$.

We will measure the radial drift due to the inclusion of viscosity by the
deviation
\begin{equation}
\delta_{\mathrm{p}}=\bar{u}^{\mathrm{V}}-\bar{u}^{\mathrm{NSH}}\;.
\end{equation}
This difference velocity $\delta_{\mathrm{p}}$ as well as the predictions of
the plate drag approximation $u_{\mathrm{PD}}$ are shown in Fig. (\ref{pd}) as
a function of the Stokes number for different turbulence parameters $\alpha$.

Fig.~\ref{pd} shows that the predictions of these two models are roughly of
the same order of magnitude if the Stokes number is about unity. For Stokes
numbers much smaller/larger than unity the results of the numerical
integration of the Navier-Stokes equations deviates from the predictions of
the plate drag model. Already Youdin and Chiang (2004) put the plate drag
approximation in question. They found that the plate drag overestimates
turbulent stresses and that vertical shear and buoyancy are important elements
missing in this description. While the plate drag model involves a radial
drift velocity which is inversely proportional to the surface density of the
layer and not explicitly dependent on particle size \cite{Wei06} found the
very contrary. In his simulations, the drift velocity shows no significant
variation with surface density, but is dependent on particle size which
clearly speaks against the validity of the plate drag model.

\subsubsection{Fitting formula}

\begin{figure}
\begin{center}
\includegraphics[scale=.45]{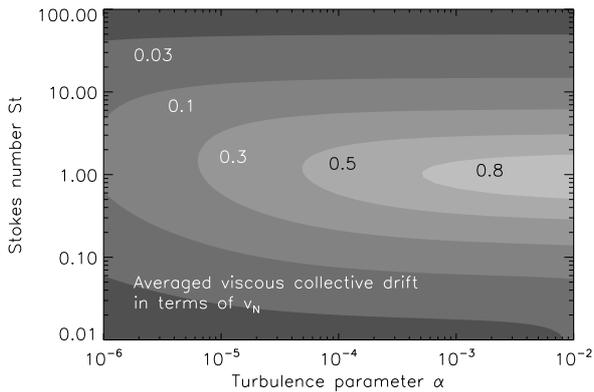}
\caption{A contour plot of the integrated radial dust velocity
$\bar{u}^{\mathrm{V}}$ under the influence of viscosity as a function of
turbulence parameter $\alpha$ and Stokes number St. The numbers related to the
contour lines on the left side, respectively, indicate the net drift velocity
in units of $-v_{\mathrm N}$.\label{resgar}}
\end{center}
\end{figure}

A simple fitting formula that reproduces the results might be useful for
forthcoming purposes for example investigations of drift time scales or radial
mixing. For this reason we fitted the vertically averaged radial dust
velocities given by the numerical solution of Eq.~(\ref{nstokes}). This
solution includes all effects investigated in this paper, i.e. collective
effects and effects of turbulent viscosity. This result is shown in
Fig.~(\ref{resgar}).

For the fit we use an expression of the form
\begin{equation}
u_{\mathrm{fit}}=\frac{\Gamma(\alpha)}{x(\alpha,\mathrm{St})+\frac{1}{x(\alpha,\mathrm{St})}}\;,
\end{equation}
following Eq.~(\ref{vipd}) for individual particle drift. In this expression
the amplitude $\Gamma$ is solely a function of the turbulence parameter
$\alpha$. The quantity $\Gamma/2$ matches the maximum occuring radial dust
velocity in units of $-v_{\mathrm{N}}$ when $\alpha$ is fixed. The parameter
$x$ is given by
\begin{equation}
x(\alpha,\mathrm{St})=10^{\xi(\alpha)}\mbox{St}^{\mu(\alpha)}\;.
\end{equation}
The two functions $\xi$ and $\mu$ are only dependent on $\alpha$. The
parameter $\xi$ determines the location of the maximum of the velocity
distribution, the parameter $\mu$ determines the width of the velocity
distribution. The fits for the three functions $\Gamma$, $\xi$ and $\mu$ were
performed with polynomials of the form
\begin{equation}
\Gamma=\sum_{\mathrm{j}=0}^{\mathrm{4}}c^{\Gamma}_{\mathrm{j}}y^{\mathrm{n}}\;,\quad
\xi=\sum_{\mathrm{j}=0}^{\mathrm{2}}c^{\xi}_{\mathrm{j}}y^{\mathrm{n}}\;,\quad
\mu=\sum_{\mathrm{j}=0}^{\mathrm{4}}c^{\mu}_{\mathrm{j}}y^{\mathrm{n}}\;,
\end{equation}
in which $y$ is given by $y=\log_{10}\alpha$. For the dependency of $\xi$ on
$\alpha$ a second order polynomial turned out to be sufficient. For the
quantities $\Gamma$ and $\mu$ a fourth order polynomial provided a satisfying
fit to the simulation results. The coefficients for all these polynomials are
listed in table (1).  The deviation between the fitting function and the
simulation within the parameter intervalls $\mbox{St}\in[10^{-2},10^2]$ and
$\alpha\in[10^{-6},10^{-2}]$ is always smaller than 0.01 $v_{\mathrm{N}}$.

\begin{table}[!h]
\label{coeffs}
\begin{tabular}{|c|c|c|c|}
\hline $i$ & $c^{\Gamma}$ & $c^{\xi}$ & $c^{\mu}$\\ \hline\hline & & & \\ $0$
& $1.89082$ & $\;\;\,2.06164\times 10^{-2}$ & $\;\;\,0.57083$\\ $1$ &
$0.14763$ & $\;\;\,4.69938\times 10^{-3}$ & $-0.41644$\\ $2$ & $0.20912$ &
$-4.05442\times 10^{-3}$ & $-0.12910$\\ $3$ & $8.25120\times 10^{-2}$ & $-$ &
$-1.24036\times 10^{-2}$\\ $4$ & $7.38181\times 10^{-3}$ & $-$ &
$-4.09782\times 10^{-4}$\\ \hline
\end{tabular}
\caption{Coefficients for the polynomical fit of the simulated integrated dust
velocities.}
\end{table}

\subsubsection{Radial drift times including effects of viscosity}

While collective effects reduce the radial drift of the dust the additional
inclusion of viscosity into the disk model again increases it. For this
reason, the time scales implied by collective effects and effects of viscosity
represent an intermediate regime between the time scales of individual
particles and the time scales implied by collective effects.

\section{Other possibilities to increase the drift timescale}\label{op}

We have seen that even with the creation of very dense midplane layers for
very low $\alpha$ the radial drift is too fast to explain the observed
millimeter flux of these disks. We will now discuss other possible solutions
to this problem. We will first consider the effect of the dust-to-gas ratio on
the drift time scales. We will then investigate the importance of the
turbulence parameter $q$ and consider the possibility of non-linear effects
which could play an important role.
 
\subsection{Dust-to-gas ratio}

In this subsection we will investigate the influence of the dust-to-gas ratio
on the drift time scales. To increase this quantity we will remove a certain
fraction of the gas from the disk. This removal has important implications for
the drift time scales. When gas is removed from the disk then the dust
particles are less affected by the motions of the gas. This leads to thinner
dust layers and hence to higher dust-to-gas ratios. For this reason collective
effects become of importance which reduces the radial drift velocities
according to Eq.~(\ref{vcoll}). In this paragraph, we will investigate how
much gas we have to remove from the disk to provide time scales larger than 2
Myrs.
\begin{figure}
\begin{center}
\includegraphics[scale=.45]{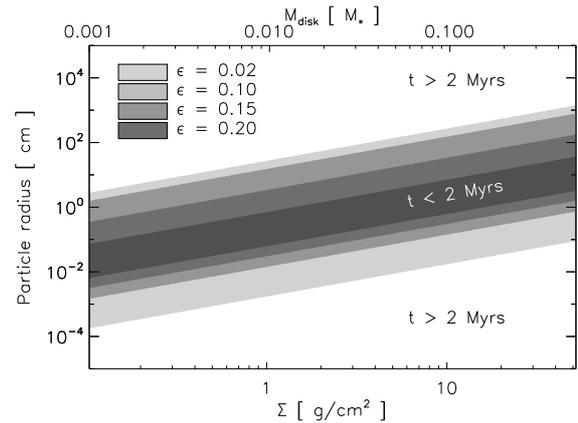}
\caption{This figure shows the effect of the dust-to-gas ratio on the drift
time scales for a turbulence parameter of $\alpha=10^{-6}$. The figure gives
the particle radius interval for which the drift time scale is smaller than 2
Myr as a function of disk mass and surface density for different total
vertical dust-to-gas ratios $\epsilon=\Sigma_d/\Sigma_g$. The surface density
is given at 100 AU. The disk mass and the surface density in this figure are
the 'original' mass and surface density of the disk before the gas depletion
that is invoked to alter the dust-to-gas ratio.
\label{ge1}}
\end{center}
\end{figure} 
\begin{figure}
\begin{center}
\includegraphics[scale=.45]{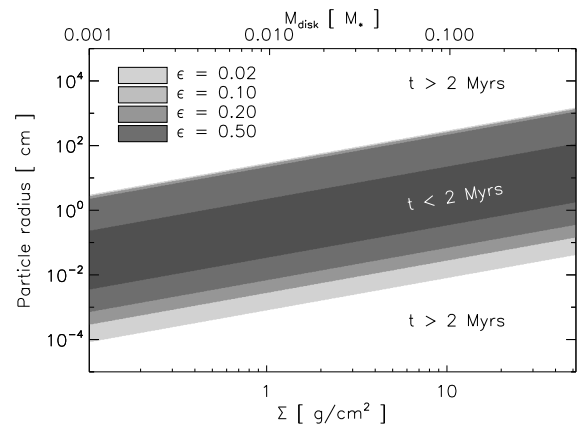}
\caption{This plot is the same as Fig.~\ref{ge1}. But here the turbulent
  $\alpha$ parameter is $10^{-5}$.\label{ge2}}
\end{center}
\end{figure}

As in the last sections we calculate the particle radius interval in which the
drift time scale is shorter than 2 Myr. These calculations were performed for
two different turbulence parameters $\alpha=10^{-5}$ and $\alpha=10^{-6}$. The
results of these simulations are shown in Fig.~(\ref{ge1})-(\ref{ge2}).  In
these figures the disk mass and the surface density are the 'original' mass
and surface density of the disk before the assumed gas depletion that is
invoked to alter the dust-to-gas ratio.

The drift behaviour of the dust particles hardly changes if only a small
fraction of the gas is removed. Both figures show that the critical particle
radius interval is little affected by removing 50\% of the gas, cf.\
Fig.~\ref{coll1}. However, for higher dust-to-gas ratios the critical interval
decreases continuously. Considering the case $\alpha=10^{-6}$ we find that cm
particles are able to stay in the outer parts of the disk for 'original' disk
masses $<0.05M_{\star}$ and $>0.2M_{\star}$ if only 5\% of the gas is
left. The critical interval disappears completely if the total vertical
dust-to-gas ratio $\epsilon=\Sigma_d/\Sigma_g$ exceeds 0.40. For higher
turbulence parameters the critical radius interval decreases slower with
higher dust-to-gas ratios. We find that for $\alpha=10^{-5}$ the interval
disappears for a dust-to-gas ratio of $\epsilon=0.75$. We conclude that
removing the gas may be a possibility to preserve mm to cm particles in the
outer part of the disk.

\subsection{Turbulence parameter $q$}

Little attention was given to the turbulence parameter $q$ until now
(cf. Appendix~\ref{apres}). A certain diffusion coefficient of the gas may be
realized by big gas eddies moving slow or by small gas eddies moving
fast. These two extreme cases are represented by $q=1$ and $q=0$,
respectively. To illustrate how strongly $q$ may influence the thickness of
the dust layer we consider the following numerical example. We assume a Stokes
number of unity and a turbulent $\alpha$ parameter of $10^{-3}$. For the
extreme case $q=0$ we calculate a dust scale height of $h/H=10^{-3}$ and for
$q=1$ we obtain $h/H=3\times 10^{-2}$. These two dust scale heights differ by
a factor of 30 which possibly influences the drift time scales.
\begin{figure}
\begin{center}
\includegraphics[scale=.45]{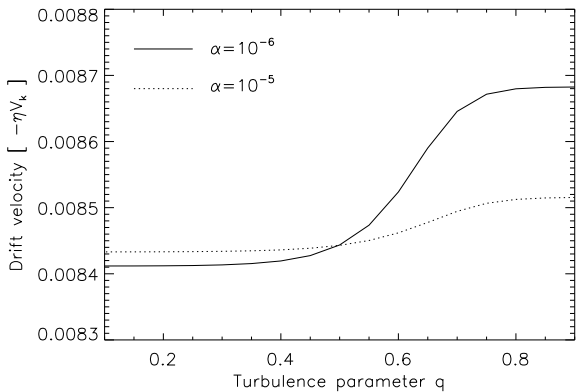}
\caption{This figure shows the effect of the turbulence parameter $q$ on the
radial drift velocity for two different turbulence parameters $\alpha$. The
Stokes number was chosen in a way that the radial drift velocity for $q=1/2$
corresponds to a timescale of 1 Myrs with regard to 100 AU assuming that
$v_{\mathrm{N}}=60$ m/s.\label{q} For $\alpha=10^{-6}$ and $\alpha=10^{-5}$
this implies $\mathrm{St}=225$ and $\mathrm{St}=235$, respectively.}
\end{center}
\end{figure}

We calculated the effect of the turbulence parameter $q$ on the drift velocity
for two different turbulence parameters $\alpha$, i.e. $\alpha=10^{-6}$ and
$\alpha=10^{-5}$. The Stokes number was chosen in a way the the drift velocity
for $q=1/2$ corresponds to a timescale of 1 Myrs. For $\alpha=10^{-6}$ and
$\alpha=10^{-5}$ this implies $\mathrm{St}=225$ and $\mathrm{St}=235$,
respectively. The results of these simulations are shown in Fig.~\ref{q}.

This figure shows that the integrated drift velocities of the dust vary by 3\%
for $\alpha=10^{-6}$ and by 1\% for $\alpha=10^{-5}$ when $q$ is changed from
zero to unity. We find that for higher $\alpha$ parameters this variation is
always less than 0.4\%. We conclude that $q$ has a very minor effect on the
radial drift velocities.

A small $h/H$ ratio leads to a high dust-to-gas ratio in the midplane of the
disk. This causes the radial drift of the dust in the midplane to decrease due
to collective effects. One would now intuitively think that a continuously
decreasing $h/H$ ratio leads to smaller and smaller radial drift velocities
but this is not the case for the following reason: When the ratio $h/H$
decreases the vertical gradients of the azimuthal gas and dust velocities
increase. Therefore, more angular momentum is transported in the higher
regions of the disk. The midplane loses angular momentum which directly causes
the radial velocity of the dust to increase. 

Finally both effects, the decrease in radial velocity due to collective
effects and the increase of velocity due to the vertical transport of angular
momentum, seem to cancel each other (or at most result in a negligible change
in radial velocity) when small $h/H$ ratios are considered.

\subsection{Effects of non-linear dynamics}

So far we have considered different equilibrium states that would potentially
allow solid particles to reside at large orbital radii for a longer time than a
single test particle. It is however also possible for dynamical effects, such
as spiral arms, turbulence or vortices, to reduce the radial drift.

Dust particles are forced to climb up the local gas pressure gradient. In the
simple case of a gas pressure that falls monotonically with radius the
particles fall into the inner disc, but the particles may end up in any local
gas overdensity that they encounter on their way \citep{klalin01,HagBos03}. If
the disc is massive enough to be gravitationally unstable, its spiral arms may
act as such local density maxima \citep{Rice04}. Transient overdensities that
occur in magnetorotational turbulence, in a way very elongated vortices, have
the same effect \citep{JohKlaHen06}, slowing down radial drift by a factor of
two. The important parameter for reducing the overall radial drift is the
life-time of the gas overdensity.  Spiral arms would from this perspective be
a good candidate, since turbulent overdensities tend to live no longer than a
few local orbits of the disc.

The coupled flow of gas and dust is in itself unstable to the streaming
instability \citep{yougoo05}, leading to particle clumping in the non-linear
state \citep{JohKlaHen06}. These local dust overdensities drag the gas along
with their orbital motion, thus reducing the sub-Keplerian head wind and the
radial drift. The effect of the streaming instability on the radial drift can
be as high as a factor two in reduction \citep{johyou07}.

\subsection{Temperature and surface density profiles}
We also investigated to which extent the radial drift time scales of the dust
particles depend on the temperature and the surface density profile. These
quantities were found to play a minor role which can be reasoned as
follows. If the temperature is decreased by a factor of 2 then the maximum
radial drift velocity $v_{\mathrm{N}}$ decreases by the same factor
(cf. Eq.~\ref{etavkk}). This means that if the temperature at 100~AU is
decreased from 20~K to a rather low value of 10~K then the drift time scales
would only increase by a factor of 2. Since the radial drift problem spans at
least one order of magnitude a change in temperature does not provide a
solution. A change in the surface density index from $\delta=0.8$ towards
$\delta=0$ changes the maximum drift velocity $v_{\mathrm{N}}$ by a factor of
1.5 according to Eq.~(\ref{etavkk}) which also does not solve the problem.  

\section{Summary}

We discussed two different subjects. First, we investigated the velocity
structure of the dust layer and the gas layer. We calculated typical radial
drift velocities of the dust including several effects, i.e. collective
effects and effects due to turbulent viscosity. The results of this first
issue enabled us to estimate characteristic drift time scales of the dust
which represents the second area of interest. These calculations were
performed in order to explain interferometric millimeter observations that
indicate the presence of mm to cm size particles in the outer part of disks
with ages up to 10 Myrs. In the following we will briefly summarize the
results.

\subsection{Velocity structure of gas and dust layer}

First we discussed the individual particle drift of the dust
\citep{Whipple72,Wei77}. In this model the dust particles are considered to be
independent and the gas is assumed not to be affected by the dust at all. This
scenario involves no vertical structure of the dust since only individual dust
particles are considered. 

We discussed the influence of collective behaviour on the radial drift
velocities of the dust. Collective effects take place when the dust-to-gas
ratio exceeds unity, e.g. when the dust sediments into a thin midplane layer
due to low turbulence. By assuming an initial dust-to-gas ratio of 0.01 we
found that the local radial drift velocity of the dust in the midplane of the
disk may be reduced by a factor of 100 for small turbulence parameters and
certain particle characterictics. The vertically averaged radial dust velocity
is at least an order of magnitude lower than the corresponding individual
particle drift.

The additional inclusion of turbulent viscosity in our considerations
increased the radial drift of the dust due to the vertical transport of
angular momentum. This increase in radial velocity was not as significant as
the decrease by including collective effects. Turbulent viscosity enhanced the
radial drift by a factor of 2 at most. Therefore, we conclude that collective
effects play the major role in the parameter regimes considered in this paper.

It turned out to be neccessary to include radial viscosity terms of gas and
dust. If these terms are neglected, the numerical solutions showed very steep
and unrealistic velocity gradients. Radial viscosity, once included in the
considerations, reduced these velocity differences which considerably changed
the radial velocities of gas and dust.

For turbulence parameters $\alpha$ larger than $10^{-4}$ collective effects as
well as effects of turbulent viscosity play a minor role. The dust is stirred
up away from the midplane in such a way that the dust-to-gas ratio is too low
for these effects to play a role. The radial drift velocities of the dust in
this case were reduced by a factor of 1.3 at most. For this reason, both
effects can be disregarded in this regime.

We compared the results of our numerical solutions to the predictions of the
so-called 'plate drag approximation'. It was shown that both models roughly
agree if the Stokes number of the particle is about unity. For Stokes numbers
higher (lower) than unity the two models predict different radial
velocities. Previous work already put the 'plate drag approximation' into
question \citep{YouChi04} with similar conclusions.

Apart from radial drift behaviour some structural aspects of the gas layer and
dust layer were also investigated. We calculated the extent of the vertical
region in which gas and dust affect each other. This calculation showed that
the gas layer behaves like a classical Ekman layer if Stokes numbers larger
than unity are considered. For well coupled particles, i.e. for particles with
Stokes number smaller than unity, the two descriptions differ significantly.

\subsection{Radial drift time scales}

We investigated the radial drift of dust particles with respect to collective
behaviour and effects of viscosity under standart conditions, i.e. an initial
dust-to-gas ratio of $10^{-2}$, standard turbulence parameter $q$, etc. We
found that a possible solution for the drift problem are very high/low disk
masses. Moreover, high porosity particles ($f<0.1$) in low turbulent disks
with masses $>0.2$ $M_{\star}$ provided drift time scales $>2$ Myrs.

However, except for these two cases the drift time scales of the dust turned
out to be far too short to explain the millimeter observations. For this
reason we investigated several possibilities in order to increase the drift
time scales. We found that removing the gas from the disk may increase the
time scales up to more than 2 Myrs. For example, for a turbulence parameter of
$\alpha=10^{-6}$ we found that cm size particles remain in the outer parts of
the disk for more than 2 Myrs if disk masses $<0.05M_{\star}$ or
$>0.2M_{\star}$ are considered and only 5\% of the gas is left. If the disk is
more turbulent more gas has to be removed in order to keep the particles in
the outer parts for longer periods of time.

Gas might be removed in the later stages of disk evolution when
photoevaporation sets in. While the gas is evaporated from the disk the dust
particles $>20$ $\mu m$ \citep{TakLin05,ThrBal05} remain in the outer parts of
the disk. However, current photoevaporation models remove the gas from the
disk rather abruptly \citep{acp06}. This would lead to rather high relative
particle velocities in the disk. Hence, the dust particles would fragment and
destructive collisions would play an important role. The centimeter particles
would be destroyed in short time scales which is not in agreement with the
observations.

We also investigated the effect of the turbulence parameter $q$ on the drift
time scales and we found that this parameter plays a minor role. Varying this
parameter between zero and unity, the most extreme cases possible, changed the
drift velocities by 3\% at most.

We did not consider the growth of the particles in this paper. The growth time
scale of the dust to reach a centimeter in size could be of the order of 1 Myr
considering the outer parts of the disk. A model including both processes,
radial drift and coagulation, would clarify this issue. Recent work about the
drift time scales in comparison to the growth time scales was done by
\cite{klabod06}. According to their calculations, the dust would long have
drifted away from 100 AU before the particles even reach the size of about a
centimeter.

Moreover, the effect of fragmentation could change the situation. Even though
a process that destroys particles is included it could finally speed up the
process of coagulation: Fragmentation leads to a permanent amount of small
particles as a result of collisions. These small particles may be swept up by
larger particles due to their high relative velocity. Although some particles
are destroyed, the final sum of both effects might lead to an accelerated
growth. These effects, the radial drift, the dust particle coagulation and the
effect of fragmentation, will be the topic of a forthcoming paper.

\subsection{Observational implications}

We have shown that thin midplane layers in low turbulent disks can conceivably
explain the presence of the observed large amounts of mm/cm sized grains if a
significant fraction of the gas is removed, for instance through
photoevaporation of the gas. Therefore, we should also discuss whether the
presence of such thin midplane layers is consistent with observations of
e.g. edge-on disks.

The infrared emission from protoplanetary disks originates from smaller
($\ll3$ mm) dust grains. These grains must be smaller than 3 micrometer, as
can be inferred from the presence of a 10 micron silicate feature in emission
in most sources. Even with relatively little vertical mixing (low turbulence)
the very smallest dust particles can still be mixed up to intermediate height
above the midplane \citep{DD04b}, although we admit that the disk should look
significantly less flared in such a case, i.e. the disk should be of Group II
in the \cite{meeus01} classification. Interestingly, \cite{acke04} have shown
that there is indeed a correlation between the presence of large grains in the
outer regions of disks around Herbig stars and the type of SED of the disk:
disks with large amounts of large (mm/cm) grains appear on average to also
have SEDs consistent with flatter disk geometry. This seems to substantiate at
least qualitatively the idea of low turbulent disks.

For larger grains, which can be observed at mm/cm wavelengths, there have not
been observations of edge-on disk with very thin mm disks so far. However,
this has two reasons. The first reason is that if the mm/cm disk is very thin,
then the chance to observe it sufficiently precisely edge-on is very
small. This reduces the number of potential candidates for such measurements
drastically. The second reason is that current state-of-the-art
interferometers do not yet have the spatial resolution to resolve such thin
disks. For instance, the Butterfly Nebula (a well-studied nearly perfectly
edge-on disk), was resolved with OVRO by \cite{wolf03}, but the vertical
extent of the observed disk was as large as the beam size, i.e. unresolved in
vertical direction.

\section{Conclusion}

We find that the radial drift time scales for millimeter-sized grains observed
in the outer ($\sim$ 100 AU) regions of protoplanetary disks are of the order
of $10^5$ years, whereas the age of these disks is generally between $10^6$
and $10^7$ years. So, according to theory, the particles that are observed
should have long ago drifted into the star, which is evidently inconsistent
with observations.  We have attempted to resolve this mystery by investigating
what happens in the case of very low-turbulence disks in which the dust
gathers in a thin midplane layer. The hope was that collective effects of the
dust (by coupling to the gas) could slow down the drift. We find that this
reduction factor is sufficient if either very high/low disk masses are
considered or the particles have high porosities. In any other case this
reduction factor is by far insufficient to resolve the problem.  Moreover, we
find that by removing a substantial fraction of the gas from the disk the
drift time scales can be sufficiently enhanced. Another possibility is
particle trapping in local gas pressure maxima \citep{KlaHen97}, but this
topic was beyond the semi-analytical model of this paper.

\section*{Acknowledgements}

We wish to thank Carsten Dominik, Dmitry Semenov, Jens Rodmann and Karim
Fertikh for useful discussions and comments. We also wish to thank the
referee, Stu Weidenschilling, for useful criticism that helped us improve the
paper.


\begin{thebibliography}{62}
\expandafter\ifx\csname natexlab\endcsname\relax\def\natexlab#1{#1}\fi

\bibitem[{{Acke} {et~al.}(2004){Acke}, {van den Ancker}, {Dullemond}, {van
  Boekel}, \& {Waters}}]{acke04}
{Acke}, B., {van den Ancker}, M.~E., {Dullemond}, C.~P., {van Boekel}, R., \&
  {Waters}, L.~B.~F.~M. 2004, \aap, 422, 621

\bibitem[{{Alexander} \& {Armitage}(2007)}]{alearm07}
{Alexander}, R.~D. \& {Armitage}, P.~J. 2007, \mnras, 375, 500

\bibitem[{{Alexander} {et~al.}(2006){Alexander}, {Clarke}, \&
  {Pringle}}]{acp06}
{Alexander}, R.~D., {Clarke}, C.~J., \& {Pringle}, J.~E. 2006, \mnras, 369, 229

\bibitem[{{Balbus} \& {Hawley}(1991)}]{balhaw91}
{Balbus}, S.~A. \& {Hawley}, J.~F. 1991, \apj, 376, 214

\bibitem[{{Balbus} \& {Hawley}(1998)}]{balhaw98}
{Balbus}, S.~A. \& {Hawley}, J.~F. 1998, Reviews of Modern Physics, 70, 1

\bibitem[{{Bally} {et~al.}(2005){Bally}, {Moeckel}, \& {Throop}}]{ThrBal05}
{Bally}, J., {Moeckel}, N., \& {Throop}, H. 2005, in ASP Conf. Ser. 341:
  Chondrites and the Protoplanetary Disk, ed. A.~N. {Krot}, E.~R.~D. {Scott},
  \& B.~{Reipurth}, 81--+

\bibitem[{{Barge} \& {Sommeria}(1995)}]{barsom95}
{Barge}, P. \& {Sommeria}, J. 1995, \aap, 295, L1

\bibitem[{{Beckwith} {et~al.}(2000){Beckwith}, {Henning}, \&
  {Nakagawa}}]{beck00}
{Beckwith}, S.~V.~W., {Henning}, T., \& {Nakagawa}, Y. 2000, Protostars and
  Planets IV, 533

\bibitem[{{Blum} \& {Wurm}(2000)}]{BluWur99}
{Blum}, J. \& {Wurm}, G. 2000, Icarus, 143, 138

\bibitem[{{Brandenburg} {et~al.}(1995){Brandenburg}, {Nordlund}, {Stein}, \&
  {Torkelsson}}]{Brand95}
{Brandenburg}, A., {Nordlund}, A., {Stein}, R.~F., \& {Torkelsson}, U. 1995,
  \apj, 446, 741

\bibitem[{{Chandrasekhar}(1961)}]{Cha61}
{Chandrasekhar}, S. 1961, {Hydrodynamic and hydromagnetic stability}
  (International Series of Monographs on Physics, Oxford: Clarendon, 1961)

\bibitem[{{Cuzzi} {et~al.}(1993){Cuzzi}, {Dobrovolskis}, \&
  {Champney}}]{CuzDobCha93}
{Cuzzi}, J.~N., {Dobrovolskis}, A.~R., \& {Champney}, J.~M. 1993, Icarus, 106,
  102

\bibitem[{{Cuzzi} \& {Weidenschilling}(2006)}]{dg}
{Cuzzi}, J.~N. \& {Weidenschilling}, S.~J. 2006, {Particle-Gas Dynamics and
  Primary Accretion} (Meteorites and the Early Solar System II), 353--381

\bibitem[{{Dominik} {et~al.}(2007){Dominik}, {Blum}, {Cuzzi}, \&
  {Wurm}}]{domppv07}
{Dominik}, C., {Blum}, J., {Cuzzi}, J.~N., \& {Wurm}, G. 2007, in Protostars
  and Planets V, ed. B.~{Reipurth}, D.~{Jewitt}, \& K.~{Keil}, 783--800

\bibitem[{{Dubrulle} {et~al.}(1995){Dubrulle}, {Morfill}, \&
  {Sterzik}}]{DubMorSte95}
{Dubrulle}, B., {Morfill}, G., \& {Sterzik}, M. 1995, Icarus, 114, 237

\bibitem[{{Dullemond} \& {Dominik}(2004{\natexlab{a}})}]{DD04b}
{Dullemond}, C.~P. \& {Dominik}, C. 2004{\natexlab{a}}, \aap, 417, 159

\bibitem[{{Dullemond} \& {Dominik}(2004{\natexlab{b}})}]{DD04}
{Dullemond}, C.~P. \& {Dominik}, C. 2004{\natexlab{b}}, \aap, 421, 1075

\bibitem[{{Fromang} \& {Nelson}(2005)}]{fronel05}
{Fromang}, S. \& {Nelson}, R.~P. 2005, \mnras, 364, L81

\bibitem[{{Garaud} {et~al.}(2004){Garaud}, {Barri{\`e}re-Fouchet}, \&
  {Lin}}]{GarFouLin04}
{Garaud}, P., {Barri{\`e}re-Fouchet}, L., \& {Lin}, D.~N.~C. 2004, \apj, 603,
  292

\bibitem[{{Goldreich} \& {Ward}(1973)}]{GolWar73}
{Goldreich}, P. \& {Ward}, W.~R. 1973, \apj, 183, 1051

\bibitem[{{G{\'o}mez} \& {Ostriker}(2005)}]{gomost05}
{G{\'o}mez}, G.~C. \& {Ostriker}, E.~C. 2005, \apj, 630, 1093

\bibitem[{{Haghighipour} \& {Boss}(2003)}]{HagBos03}
{Haghighipour}, N. \& {Boss}, A.~P. 2003, \apj, 583, 996

\bibitem[{{Hartmann} {et~al.}(1998){Hartmann}, {Calvet}, {Gullbring}, \&
  {D'Alessio}}]{Hart98}
{Hartmann}, L., {Calvet}, N., {Gullbring}, E., \& {D'Alessio}, P. 1998, \apj,
  495, 385

\bibitem[{{Johansen} {et~al.}(2006{\natexlab{a}}){Johansen}, {Henning}, \&
  {Klahr}}]{jhk06}
{Johansen}, A., {Henning}, T., \& {Klahr}, H. 2006{\natexlab{a}}, \apj, 643,
  1219

\bibitem[{{Johansen} \& {Klahr}(2005)}]{Johans05}
{Johansen}, A. \& {Klahr}, H. 2005, \apj, 634, 1353

\bibitem[{{Johansen} {et~al.}(2006{\natexlab{b}}){Johansen}, {Klahr}, \&
  {Henning}}]{JohKlaHen06}
{Johansen}, A., {Klahr}, H., \& {Henning}, T. 2006{\natexlab{b}}, \apj, 636,
  1121

\bibitem[{{Johansen} \& {Youdin}(2007)}]{johyou07}
{Johansen}, A. \& {Youdin}, A. 2007, ArXiv Astrophysics e-prints

\bibitem[{{Kempf} {et~al.}(2000){Kempf}, {Blum}, \& {Wurm}}]{Kem00}
{Kempf}, S., {Blum}, J., \& {Wurm}, G. 2000, Bulletin of the American
  Astronomical Society, 32, 1099

\bibitem[{{Kitamura} {et~al.}(2002){Kitamura}, {Momose}, {Yokogawa}, {Kawabe},
  {Tamura}, \& {Ida}}]{Kit02}
{Kitamura}, Y., {Momose}, M., {Yokogawa}, S., {et~al.} 2002, \apj, 581, 357

\bibitem[{{Klahr} \& {Bodenheimer}(2006)}]{klabod06}
{Klahr}, H. \& {Bodenheimer}, P. 2006, \apj, 639, 432

\bibitem[{{Klahr} \& {Henning}(1997)}]{KlaHen97}
{Klahr}, H.~H. \& {Henning}, T. 1997, Icarus, 128, 213

\bibitem[{{Klahr} \& {Lin}(2001)}]{klalin01}
{Klahr}, H.~H. \& {Lin}, D.~N.~C. 2001, \apj, 554, 1095

\bibitem[{{Launder}(1976)}]{lau76}
{Launder}, B.~E. 1976, {Heat and mass transport} (Turbulence.~(A77-20355 07-34)
  Berlin and New York, Springer-Verlag, 1976, p.~231-287.), 231--287

\bibitem[{{McComb}(1990)}]{mcc90}
{McComb}, W.~D. 1990, Chemical Physics

\bibitem[{{Meeus} {et~al.}(2001){Meeus}, {Waters}, {Bouwman}, {van den Ancker},
  {Waelkens}, \& {Malfait}}]{meeus01}
{Meeus}, G., {Waters}, L.~B.~F.~M., {Bouwman}, J., {et~al.} 2001, \aap, 365,
  476

\bibitem[{{Nakagawa} {et~al.}(1986){Nakagawa}, {Sekiya}, \&
  {Hayashi}}]{NakSekHay86}
{Nakagawa}, Y., {Sekiya}, M., \& {Hayashi}, C. 1986, Icarus, 67, 375

\bibitem[{{Natta}(2004)}]{natta04}
{Natta}, A. 2004, in ASP Conf. Ser. 324: Debris Disks and the Formation of
  Planets, ed. L.~{Caroff}, L.~J. {Moon}, D.~{Backman}, \& E.~{Praton}, 20--+

\bibitem[{{Natta} {et~al.}(2007){Natta}, {Testi}, {Calvet}, {Henning},
  {Waters}, \& {Wilner}}]{natta06}
{Natta}, A., {Testi}, L., {Calvet}, N., {et~al.} 2007, in Protostars and
  Planets V, ed. B.~{Reipurth}, D.~{Jewitt}, \& K.~{Keil}, 767--781

\bibitem[{{Natta} {et~al.}(2004){Natta}, {Testi}, {Neri}, {Shepherd}, \&
  {Wilner}}]{nattesner04}
{Natta}, A., {Testi}, L., {Neri}, R., {Shepherd}, D.~S., \& {Wilner}, D.~J.
  2004, \aap, 416, 179

\bibitem[{{Ormel} {et~al.}(2007){Ormel}, {Spaans}, \& {Tielens}}]{ormel07}
{Ormel}, C.~W., {Spaans}, M., \& {Tielens}, A.~G.~G.~M. 2007, \aap, 461, 215

\bibitem[{{Rice} {et~al.}(2004){Rice}, {Lodato}, {Pringle}, {Armitage}, \&
  {Bonnell}}]{Rice04}
{Rice}, W.~K.~M., {Lodato}, G., {Pringle}, J.~E., {Armitage}, P.~J., \&
  {Bonnell}, I.~A. 2004, \mnras, 355, 543

\bibitem[{{Rodmann}(2006)}]{rodphd}
{Rodmann}, J. 2006, PhD thesis, PhD Thesis, Combined Faculties for the Natural
  Sciences and for Mathematics of the University of Heidelberg,
  Germany.~XIII+137 pp.~(2006)

\bibitem[{{Rodmann} {et~al.}(2006){Rodmann}, {Henning}, {Chandler}, {Mundy}, \&
  {Wilner}}]{Rod06}
{Rodmann}, J., {Henning}, T., {Chandler}, C.~J., {Mundy}, L.~G., \& {Wilner},
  D.~J. 2006, \aap, 446, 211

\bibitem[{{Schr{\"a}pler} \& {Henning}(2004)}]{schr04}
{Schr{\"a}pler}, R. \& {Henning}, T. 2004, \apj, 614, 960

\bibitem[{{Sekiya}(1998)}]{sek98}
{Sekiya}, M. 1998, Icarus, 133, 298

\bibitem[{{Shakura} \& {Sunyaev}(1973)}]{ShaSun73}
{Shakura}, N.~I. \& {Sunyaev}, R.~A. 1973, \aap, 24, 337

\bibitem[{{Suttner} \& {Yorke}(2001)}]{SutYor01}
{Suttner}, G. \& {Yorke}, H.~W. 2001, \apj, 551, 461

\bibitem[{{Takeuchi} \& {Lin}(2002)}]{TakLin02}
{Takeuchi}, T. \& {Lin}, D.~N.~C. 2002, \apj, 581, 1344

\bibitem[{{Takeuchi} \& {Lin}(2005)}]{TakLin05}
{Takeuchi}, T. \& {Lin}, D.~N.~C. 2005, \apj, 623, 482

\bibitem[{{Testi} {et~al.}(2003){Testi}, {Natta}, {Shepherd}, \&
  {Wilner}}]{Tes03}
{Testi}, L., {Natta}, A., {Shepherd}, D.~S., \& {Wilner}, D.~J. 2003, \aap,
  403, 323

\bibitem[{{V\"olk} {et~al.}(1980){V\"olk}, {Morfill}, {Roeser}, \&
  {Jones}}]{voelk80}
{V\"olk}, H.~J., {Morfill}, G.~E., {Roeser}, S., \& {Jones}, F.~C. 1980, \aap,
  85, 316

\bibitem[{{Weidenschilling}(1977)}]{Wei77}
{Weidenschilling}, S.~J. 1977, \mnras, 180, 57

\bibitem[{{Weidenschilling}(1979)}]{Wei79}
{Weidenschilling}, S.~J. 1979, Bulletin of the American Astronomical Society,
  11, 552

\bibitem[{{Weidenschilling}(1980)}]{Wei80}
{Weidenschilling}, S.~J. 1980, Icarus, 44, 172

\bibitem[{{Weidenschilling}(2003)}]{Wei03}
{Weidenschilling}, S.~J. 2003, Icarus, 165, 438

\bibitem[{{Weidenschilling}(2006)}]{Wei06}
{Weidenschilling}, S.~J. 2006, Icarus, 181, 572

\bibitem[{{Whipple}(1972)}]{Whipple72}
{Whipple}, F.~L. 1972, in From Plasma to Planet, ed. A.~{Elvius}, 211--+

\bibitem[{{Wilner} {et~al.}(2005){Wilner}, {D'Alessio}, {Calvet}, {Claussen},
  \& {Hartmann}}]{Wil05}
{Wilner}, D.~J., {D'Alessio}, P., {Calvet}, N., {Claussen}, M.~J., \&
  {Hartmann}, L. 2005, \apjl, 626, L109

\bibitem[{{Wolf} {et~al.}(2003){Wolf}, {Padgett}, \& {Stapelfeldt}}]{wolf03}
{Wolf}, S., {Padgett}, D.~L., \& {Stapelfeldt}, K.~R. 2003, \apj, 588, 373

\bibitem[{{Youdin} \& {Chiang}(2004)}]{YouChi04}
{Youdin}, A.~N. \& {Chiang}, E.~I. 2004, \apj, 601, 1109

\bibitem[{{Youdin} \& {Goodman}(2005)}]{yougoo05}
{Youdin}, A.~N. \& {Goodman}, J. 2005, \apj, 620, 459

\bibitem[{{Youdin} \& {Shu}(2002)}]{YouShu02}
{Youdin}, A.~N. \& {Shu}, F.~H. 2002, \apj, 580, 494

\end{thebibliography}

\appendix

\section{Theoretical background}\label{tb}

The main ingredients of the models are the stellar parameters, the structure
and mass of the disk, the description of the turbulence, and the description
of the interaction between the dust and the gas. In principle, for the radial
drift velocity we do not need knowledge of the turbulence (except for the
turbulence effects described by \cite{JohKlaHen06} which might reduce the
drift by a factor of up to 2). However, turbulence sets the thickness of the
dust midplane layer, and hence the density of the dust, which affects the
drift velocity once this density comes close to, or exceeds the gas
density. If there would be no turbulence at all, the dust particles would
settle to the midplane and form a razor-thin layer qualitatively like the
rings of Saturn. The more turbulence there is the harder it is for the dust
to form a thin layer since it is mixed up and again transported in the higher
regions of the disk. Turbulence is not yet understood in detail and there is
various literature about this challenging topic
\citep{voelk80,schr04,balhaw91,balhaw98}. In this paper we will make use of
the $\alpha$-prescription \citep{ShaSun73}. This specific way of regarding
turbulence is somewhat superficial since it ignores most details of
turbulence. However, the advantage is that it makes turbulence manageable
without extensive hydrodynamical simulations.

\subsection{$\alpha$-prescription for turbulent gas}\label{apres}

In this section we will give a short overview about our $\alpha$-prescription
of particles in a turbulent disk. Turbulence mixes things up and therefore
acts like a kind of diffusion. The diffusion coefficient\footnote{All symbols
used in this paper are listed in a table at the very end.} $D$ can be written
as a product of a velocity scale $V_0$ and a length scale $L_0$,
\begin{equation}
\label{dn}
D=V_0\times L_0\;.
\end{equation}
A typical velocity scale is the isothermal soundspeed $\cs^2\equiv
k\Tgas/\mu$, with $\Tgas$ the gas temperature, $k$ the Boltzmann constant,
$\mu$ the mean molecular weight of the gas (which we take 2.3 times the proton
mass for a standard mixture of molecular hydrogen and atomic helium). A
characteristic length scale is given by the pressure scale height of the gas
$H=c_s/\Omega_k$, where $\Omega_k \equiv \sqrt{G M_{*}/r^3}$.  Regarding these
typical scales at a certain radius we can alternatively express the diffusion
coefficient for the gas as follows
\begin{equation}
\label{dvl}
D=\alpha c_sH\;.
\end{equation}
The value of the parameter $\alpha$ reflects the amount of turbulence in the
disk and it ranges from $10^{-6}$ \citep{Wei80} up to $10^{-2}$
\citep{Hart98}. Extensive hydrodynamical simulations show typical $\alpha$
values of about $10^{-3}$ (\cite{Brand95}, \cite{Johans05}).  In general we
have to distinguish between the $\alpha$ introduced to determine the vertical
distribution of the dust and the $\alpha_{acc}$ linked with the accretion
process. Both quantities are related by $\alpha_{acc}=$Pt$\alpha$ where Pt
denotes the Schmidt number.  We will assume Pt$=1$ in this paper
\citep{lau76,mcc90,Johans05}. 
  
There is an ambiguity with regard to this formulation till now since $\alpha$
does not provide any information about $V_0$ and $L_0$. This can be seen in a
better way by introducing a turbulence parameter $q$
\begin{equation}
D=\alpha c_sH=\alpha^qc_s\times \alpha^{1-q}H\;,
\end{equation}
which ranges between 0 and 1. Now we can identify the terms of equation
(\ref{dn}) by defining $V_0=\alpha^qc_s$ and $L_0=\alpha^{1-q}H$. The physical
interpretation of $q$ is as follows. A certain diffusion can be realized by
big eddies which move very slow ($q=1$) or by small eddies moving very fast
($q=0$). Both possibilities result in the same diffusion-coefficient. Normally
$q$ is set to be 1/2 (e.g.~\cite{schr04}) but there are other values
possible. For example, in self-induced turbulence $q$ tends to be smaller than
1/2 \citep{Wei06}. Another possibility are big convective cells of scale $H$
\citep{KlaHen97} which would imply $q=1$. The fundamental importance of
knowledge about $q$ can be understood by calculating the turn-over-eddy-time
$\teddy$
\begin{equation}
\label{etime}
\teddy=\frac{L_0}{V_0}=\frac{2\pi}{\omeddy}=\alpha^{1-2q}\frac{1}{\Omega_k}\;.
\end{equation}
where $\omeddy$ is the eddy turn-over frequency.  Comparing the two extreme
cases $q=0$ and $q=1$ the timescale $\teddy$ changes by a factor of
$\alpha^2$. Taking a typical value of $\alpha=10^{-3}$ means that these
time scales differ by 6 orders of magnitude. In the next section we will see
that the effect of the turbulent gas on the dust, and therefore on the dust
scale height, is highly dependent on this timescale $\teddy$. For this reason
knowledge about $q$ is an essential requirement.

\subsection{Viscosity of gas and dust}

In this paper we will assume that the value of the turbulent viscosity of the
gas is basically the turbulent diffusion coefficient of the gas
eq.~(\ref{dvl}), which corresponds to $\mathrm{Pt}=1$. \cite{Johans05} found
that the ratio between these two quantities ranges from 0.8 up to 1.6 in MRI
turbulence, depending on the direction, i.e. vertical or radial.

We will also assume a certain turbulent viscosity of the dust. This quantity
can be expressed as the product of a characteristic mixing length and a
relative turbulent velocity of the particles
\begin{equation}
\nu_{\mathrm{d}}=L_{\mathrm{mix}}v_{\mathrm{t}}\;.
\end{equation}
For Stokes numbers greater than unity the relative turbulent velocity between
the dust particles is given by $v_{\mathrm{t}}=V_0/\mbox{St}$
\citep{voelk80,SutYor01}. The velocity $V_0$ denotes the turbulent velocity of
the largest gas eddy introduced in the first sections. With the mixing length
$L_{\mathrm{mix}}=\tau_{\mathrm{s}}v_{\mathrm{t}}$ the viscosity of the dust
is given by
\begin{equation}
\nu_{\mathrm{d}}=\frac{D}{\mbox{St}}\qquad\mbox{for}\qquad\mbox{St}>1\;,
\end{equation}
where we used the Eq.~(\ref{dvl}) and (\ref{stt}). 

Now, for Stokes numbers smaller than unity the dust particles are well coupled
to the gas, i.e. both disk components, gas and dust, behave more like one
single fluid than two different types of matter. We have already seen that the
diffusion coefficient for the dust in eq.~(\ref{ddust}) matches the diffusion
coefficient for the gas in the $\mbox{St}<1$ regime. We will assume the same
behaviour with regard to the viscosity, e.g. that the dust viscosity equals
the gas viscosity for small St. Considering this, the dust viscosity in both
regimes, $\mbox{St}>1$ and $\mbox{St}<1$, is then given by \citep{schr04}
\begin{equation}
\nu_{\mathrm{d}}=\frac{D}{1+\mbox{St}}\;.
\end{equation}

\subsection{Self-induced turbulence}

In this paper we will vary $\alpha$ between a rather high value of $0.01$ down
to a rather low value of $10^{-6}$. As has been discussed by many authors (and
will also be discussed below) we expect the following behaviour: For low
$\alpha$ the dust sediments into a very thin midplane layer. However, there is
a limit on how thin this layer can become, or in other words: how low the
level of turbulence can become. \cite{Wei79} has shown that a
shear-instability between the dust layer and the gas induces a weak, but
non-negligible level of turbulence. This is called `self-induced turbulence'.

In principle this self-induced turbulence can be described by the
($\alpha$,$q$)-formalism discussed above, as long as we allow that both
$\alpha$ and $q$ depend on height above the midplane $z$ and radial distance
to the star $r$. Determining the $\alpha$ and $q$ of this self-induced
turbulence is a complex matter, and although enormous progress has been made
in this field, until now there are still many open issues
\citep{CuzDobCha93,Wei06,jhk06}.

In this paper we will not explicitly address the issues of self-induced
turbulence: we will keep $\alpha$ and $q$ as parameters of the model (assuming
that they do not depend on $z$ or St). However, in appendix \ref{sia} we will
roughly estimate the level of self-induced turbulence to obtain a lower limit
to the $\alpha$ that we are allowed to use.

\subsection{Stokes number}\label{thestokes}

A moving particle in a gas at rest loses a significant fraction of its
momentum within a time called stopping time $\tau_s$. This time depends on the
friction between the particle and the gas. A strong friction means a small
$\tau_s$, and vice versa. The friction depends on the particle cross section
$\sigma_p=\pi a^2$ and, therefore, the particle radius\footnote{We will always
assume the particles to be spherical.} $a$, the relative velocity $v_p$ with
respect to the gas and the properties of the gas (mainly gas density $\rho_g$,
isothermal sound speed $c_s$ and molecular mean free path $l$).

For particles, with a size smaller than the molecular mean free path, the
friction can be expressed by a simple formula:
\begin{equation}
\label{epstein}
F_e=\frac{4}{3}\rho_gc_s\sigma_pv_p 
\fullstop
\end{equation}
This is the ``Epstein drag law''. In this regime the stopping time equals:
\begin{equation}
\label{epsteintau}
\tau_s=\frac{m_pv_p}{F_e}\stackrel{\mathrm{Ep.}}{=}\frac{\rho_sa}{\rho_gc_s}\;,
\end{equation}
where $m_p$ is the particle mass, which can be expressed with the particle
material density $\rho_s$ as $m_p=(4\pi/3)\rho_s a^3$. 

For particles larger than the mean free path the drag law is much more
complex. This regime is characterized by the ``Stokes drag law''. In this
paper we focus on particles that are always smaller than the mean free path
and can ignore the Stokes regime.

If the stopping time $\tau_s$ is much smaller than the turn-over-eddy time
$\teddy$, the particles are strongly coupled to the gas having the same motions
and the same behaviour with regard to diffusion.  When $\tau_s$ exceeds
$\teddy$ by far, the dust decouples from the gas and is hardly influenced by
the turbulence of the gas.

The stopping time characterizes the dynamic properties of the particle as it
moves through the disk. Therefore, we can replace all microphysical particle
properties like $a$, $\rho_s$, $m_p$ by $\tau_s$. Particles with vastly
different properties (e.g.~size), but the same $\tau_s$ behave {\em entirely}
the same.

Instead of using the stopping time $\tau_s$, an even more convenient parameter
is the so-called ``Stokes number'' $\mbox{St}_L$. It is defined by:
\begin{equation}\label{stt}
\mbox{St}_L=\frac{\tau_s}{\teddy}=\tau_s\Omega_k\alpha^{2q-1}
\fullstop
\end{equation}
The particles are strongly coupled for St$_L\ll1$ and hardly influenced by the
gas for St$_L\gg1$. For the case $q=1/2$ the Stokes number does not depend on
$\alpha$. We will now introduce the Stokes number St by
St=St$_L(q=1/2)=\tau_s\Omega_k$ and we will formulate all the particle
properties in terms of $\mbox{St}$ and $\mbox{St}_L$.

\subsection{Thickness of the dust layer}

Let us now turn to the calculation of the thickness of the dust layer. This
thickness is determined by an equilibrium between dust which settles towards
the midplane and diffusion which stirrs the dust up again \citep{DubMorSte95,
schr04, DD04}. The settling can be described by the equation of motion for a
dust particle,
\begin{equation}\label{eq-zdd}
\ddot z = -\Omega_{\mathrm{k}}^2z-\frac{1}{\tau_{\mathrm{s}}}\dot{z}\;.
\end{equation}
For $\mbox{St}> 1/2$ the particle describes a damped oscillation around the
midplane, corresponding to an orbit inclined with respect to the midplane.
For $\mbox{St}\ll 1/2$ the particle is so well bound to the gas, that the
downward motion equals to an equilibrium settling motion with magnitude:
\begin{equation}\label{vvvs}
v_{\mathrm{sett}} = -z\mathrm{St}\Omega_{\mathrm{k}}
\comma
\end{equation}
so that the second order differential equation Eq.~(\ref{eq-zdd}) reduces
to a first order differential equation:
\begin{equation}
\dot{z}=-v_{\mathrm{sett}}
\fullstop
\end{equation}
The diffusion of the dust is characterized by
\begin{equation}
\label{ddust}
D_d=\frac{D}{1+\mbox{St}_L}\;,
\end{equation} 
which is the diffusion coefficient of the gas corrected by a factor including
the coupling between dust and gas.  This expression decreases with higher
Stokes number since the dust is not longer affected by the gas. From the
numbers $D_d$ and $\tsett$, representing settling and diffusion, we can
construct a length scale by $h^2=D_d \tsett$. 
This leads to the expression for the dust layer thickness:
\begin{equation}\label{eq-hsquare}
h^2=D_d\;\mbox{max}(\tsett,1/2\Omega_{\mathrm{k}})
\end{equation}
We resticted the settling time scale to be at least half an orbital time
scale. For St larger than unity the motion of a dust particle above the
midplane corresponds to a damped inclined orbit. The velocity towards the
midplane of the disk can not be higher than the projected Kepler velocity
$V_{\mathrm{k}}z/r$ and, hence, the time for the particle to cross the
midplane not significantly smaller than an orbital time scale. However, the
dynamics of bodies in quasi-Keplerian orbits in turbulence may be not well
described by diffusion, and the authors are aware that the approach
Eq.~(\ref{eq-hsquare}) should be used with caution. Also the vertical
distribution of the dust for the case of inclined orbits (i.e. for large
Stokes number) is not Gaussian but rather bimodal, but we will ignore this
effect. With Eqs.~(\ref{dvl}, \ref{ddust}), as well as the expression for the
gas scale height $H= c_{\mathrm{s}}/\Omega_{\mathrm{k}}$, one can rewrite
Eq.~(\ref{eq-hsquare}) as:
\begin{equation}
\label{gp}
\left(\frac{h}{H}\right)^2=\frac{\alpha}{\mbox{min}(\mathrm{St},1/2)\;(1+{\mathrm{St}}_L)}\;.
\end{equation}

This is the most general description of the dust layer. The Stokes numbers St
and St$_L$ allow us to calculate the thickness of the dust layer and,
therefore, the dust mass densities. With these densities we are able to
calculate radial drift velocities to estimate particle radial drift times in
the outer parts of the disk.
 
\subsection{Gas and dust density}\label{dismmodel}

With the results of the last section we now present the mass densities of gas
and dust. We assume the disk to be isothermal in the $z$-direction and we use
a thin disk approximation: $z\ll r$. The vertical distribution of the gas is
then given by
\begin{equation}\label{rhogas_}
\rho_{\mathrm{g}}=\frac{\Sigma_{\mathrm{g}}}{\sqrt{2\pi}H}\exp\left(-\frac{1}{2}\frac{z^2}{H^2}\right).
\end{equation}
The scale height of the gas $H$ is given by
$H=c_{\mathrm{s}}/\Omega_{\mathrm{k}}$. The density distribution of the dust
is
\begin{equation}\label{rhodust}
\rho_{\mathrm{d}}=\frac{\Sigma_{\mathrm{d}}}{\sqrt{2\pi}h}
\exp\left(-\frac{1}{2}\frac{z^2}{h^2}\right),
\end{equation}
wherein $h$ denotes the scale height of the dust we estimated in the last
section and $\Sigma_{\mathrm{d}}$ ist given by
$\epsilon_0\Sigma_{\mathrm{g}}$. The quantity $\epsilon_0$ in this expression
is the initial dust-to-gas mass ratio and it is usually set to 0.01.  Further
investigations will show that the drift time scales depend strongly on this
parameter which means that $\epsilon_0$ is one of the main quantities
describing our system.  Let us assume that the surface density
$\Sigma_{\mathrm{g}}$ of the gas follows a power law
\begin{equation}\label{surfgas}
\Sigma_{\mathrm{g}}(r)=\Sigma_0\left(\frac{r_0}{r}\right)^{\delta}\;,
\end{equation}
and the temperature $T$ does so as well:
\begin{equation}
T(r)=T_0\left(\frac{r_0}{r}\right)^{\xi}\;.
\end{equation}
We choose the power law index $\delta$ to be 0.8 following
\citet{Kit02}. Moreover, we assume the temperature $T_0$ to be 200~K,
corresponding to a passive disk irradiated under an angle of 0.05 degrees
around a T Tauri star with a surface temperature of 4000 K and $R_{\star}=2.5$
$R_{\odot}$. The power law index of the temperature $\xi$ is set to be
1/2. The mass of the disk is given by
\begin{equation}\label{eq-diskmass}
M_{\mathrm{disk}}=\int_{r_{\mathrm{ev}}}^{r_{\mathrm{max}}}\Sigma_{\mathrm{g}}(r)rdrd\varphi=\frac{2\pi\Sigma_0r_0^{\delta}}{2-\delta}\left[r_{\mathrm{max}}^{2-\delta}-r_{\mathrm{ev}}^{2-\delta}\right]\;,
\end{equation}
which we take as a parameter of our models. The size of the disk
$r_{\mathrm{max}}$ is set to be 150 AU following the observational results of
the Taurus-Auriga region by \cite{Rod06}. The dust evaporation radius
$r_{\mathrm{ev}}$ is 0.03 AU. Throughout the paper we will give the mass of
the disk in terms of the central stellar mass $M_{\star}$ which we assume to
be 1 $M_{\sun}$.

If we consider Herbig Ae/Be stars instead of T Tauri stars then the disk mass
and the outer disk radius increases \citep{natta04}. Herbig stars are usually
a factor of 4 heavier and have outer disk radii of 300 AU and more. However,
Eq.~(\ref{eq-diskmass}) shows that the surface density $\Sigma_0$ is hardly
affected by this change. The disk is more massive, but since the disk extent
increases as well the actual amount of mass per unit area in the disk does not
change.

In this paper we use a standard velocity $v_{\mathrm{N}}$. This quantity is
the velocity by which the gas moves azimuthally slower than Keplerian velocity
$V_{\mathrm{k}}$, i.e. $v_{\mathrm{gas}}=V_{\mathrm{k}}-v_{\mathrm{N}}$. It is
given by \citep{Wei77,NakSekHay86}
\begin{equation}\label{etavkk}
v_{\mathrm{N}}=-\frac{\partial_{r}p_{\mathrm{g}}}{2\rho_{\mathrm{g}}\Omega_{\mathrm{k}}}=\frac{c_{\mathrm{s}}^2}{2V_{\mathrm{k}}}\left(\frac{3}{2}+\frac{\xi}{2}+\delta\right)\;.
\end{equation}

\section{Self-induced turbulence}\label{sia}

There is a limit of how thin the dust layer can be, i.e. how low the level of
turbulence can become. The calculations in this paper are physically realistic
as long as the turbulence parameter $\alpha$ does not drop below this
value. In this part of the appendix we present the lowest possible amount of
turbulence in terms of the turbulence parameterisation $\alpha$ and $q$.

The reason for a minimal amount of turbulence is the following: when the dust
settles towards the midplane and the dust-to-gas ratio exceeds unity the gas
is carried along with the dust. Since the gas above the midplane still moves
somewhat sub-keplerian the resulting shear can destabilize the layer causing
turbulence. This instability is called Kelvin-Helmholtz instability. The
turbulence in turn again diffuses the particles until a steady state is
reached.

In the next two subsections we will translate this effect into the
$\alpha$-prescription. To do this, we have to consider two different
cases. The Stokes number $\mbox{St}_c$ which divides these two cases is
roughly given by $10^{-2}$ \citep{dg}. Let us first focus on the
$\mbox{St}<\mbox{St}_c$ case.

\subsection{$\mbox{St}<\mbox{St}_c$}

In this regime of small particles the thickness of the layer is almost
independent of particle properties and given by \citep{sek98}
\begin{eqnarray}
\label{sekiya}
\frac{h}{H}=\sqrt{\mbox{Ri}}\frac{H}{R}\;.
\end{eqnarray}
The quantity Ri is the critical Richardson number. This number, which is
usually assumed to be constant, is approximately given by $\mathrm{Ri}=0.25$
\citep{Cha61}. Recent publications call this assumption in question
\citep{gomost05} and show that the Richardson-number required for marginal
instability is eventually higher than the traditional value of 0.25 depending
on the Stokes number. However, since the results of our simulations turned out
to be hardly dependend on Ri this effect may be neglected in the context of
radial drift velocities. A comparison between (\ref{sekiya}) and (\ref{gp})
yields the turbulent $\alpha$ parameter for this case.
\begin{eqnarray}
\label{alse}
\alpha=\beta\mbox{Ri}\mbox{St}(1+\mbox{St}_L)
\end{eqnarray}
The quantity $\beta$ is simply the squared ratio between the isothermal sound
speed $c_{\mathrm{s}}$ and the Kepler velocity $V_{\mathrm{k}}$. At 100 AU the
quantity $\beta$ is roughly of the order of $10^{-2}$. This leads to a
turbulent $\alpha$ value which is at most of the order of $\alpha\sim 10^{-5}$
assuming $q=1/2$ and $\mathrm{Ri}=1/4$.

\subsection{$\mbox{St}>\mbox{St}_c$}

In this second regime the diffusion coefficient in self-induced turbulence for
the gas is given by $D=(\beta V_{\mathrm{k}})^2/\Omega_{\mathrm{
k}}\mbox{Ro}^2$ \citep{dg}. A direct comparison between the last
equation and Eq.~(\ref{dvl}) shows that the turbulence parameter is a
constant:
\begin{eqnarray}
\label{aldu}
\alpha=\frac{\beta}{\mbox{Ro}^2}\;.
\end{eqnarray}
To estimate the $q$ parameter we have to focus on the so-called Rossby number
Ro that represents the ratio between the turn-over-frequency of the largest
eddie $\omeddy$ and the Kepler-frequency, so that $\omeddy=\Omega_k$Ro. With
this relation and equation (\ref{etime}) the turbulence parameter $q$ can be
calculated to be
\begin{eqnarray}
q=\frac{1}{2}\left[1-\log_{\alpha}\left(\frac{2\pi}{\mbox{Ro}}\right)\right]\;.
\end{eqnarray}
Taking Ro=25 we get $q=0.45$ which is quite close to the value 0.5. If we
assume a $\beta$ of about $10^{-2}$ at 100~AU then the self-induced turbulent
$\alpha$ parameter in this case is roughly given by $10^{-5}$. According to
\cite{dg}, the value of the Rossby number is uncertain up to a factor of
2-3. This means that the $\alpha$-value could be even lower and also in the
order of $\sim 10^{-6}$ at 100~AU in the disk. This might be also the case if
higher stellar masses or lower temperatures are considered.

\begin{table*}[!p]
\begin{center}
\begin{tabular}{lr}
Variable & Explanation\\ \hline \hline $a$ & radius of the dust particle\\
$m_{\mathrm{p}}$ & mass of the dust particle\\ $\sigma_{\mathrm{p}}$ &
geometrical cross section of the dust particle\\ $v_{\mathrm{p}}$ & velocity
of the dust particle\\ $f$ & filling factor of the dust particle\\ \hline $r$
& distance to the central star from a point in the midplane\\ $z$ & height
above the midplane\\ $c_{\mathrm{s}}$ & isothermal soundspeed\\
$\Omega_{\mathrm{k}}$, $V_{\mathrm{k}}$ & Kepler frequency, Kepler velocity\\
$H=c_{\mathrm{s}}/\Omega_{\mathrm{k}}$ & gas scale height\\ $h$ & dust scale
height\\ $\Sigma_{\mathrm{g}}$, $\Sigma_{\mathrm{d}}$&surface density of the
gas and the dust\\ $\rho_{\mathrm{g}}$, $\rho_{\mathrm{d}}$ & gas and dust
density\\ $p_{\mathrm{g}}=c_{\mathrm{s}}^2\rho_{\mathrm{g}}$&gas pressure\\
$\epsilon_0$, $\epsilon$& initial and current dust-to-gas ratio\\
$\psi=1/(1+\epsilon)$& reduction parameter in the NSH solution\\
$\eta=-\partial_{\mathrm{r}}p_{\mathrm{g}}/2\rho_{\mathrm{g}}r\Omega_{\mathrm{k}}^2$&
ratio between radial pressure force and gravitational force\\
$r_{\mathrm{ev}}$&evaporation radius\\ $r_{\mathrm{max}}$&outer radius of the
disk\\ $M_{\star}$&mass of the central star\\ $M_{\mathrm{disk}}$&mass of the
disk\\ $v_{\mathrm{N}}=\eta V_{\mathrm{k}}$&typical radial drift scale\\
\hline $\alpha$, $q$&turbulence parameter\\ $\tau_{\mathrm{s}}$&stopping time
of the dust particle\\ $\mbox{St}=\tau_{\mathrm{s}}\Omega_{\mathrm{k}}$&
Stokes number of the particle\\ $\mbox{St}_L=\mbox{St}\alpha^{2q-1}$&modified
Stokes number\\ $L_0$, $V_0$, $\omeddy$, $\teddy$&length, velocity, frequency
scale and time of the turbulent gas motions\\ $D=\nu_{\mathrm{g}}$,
$D_{\mathrm{d}}=\nu_{\mathrm{d}}$& diffusion coefficients/viscosity of gas and
dust\\ $\mbox{Re}_{\star}$&critical Reynolds number\\ \hline $w_r$,
$w_{\varphi}$& radial and azimuthal velocity in the single particle drift
equations\\ $u^{\mathrm{NSH}}$&radial velocity of the dust in the NSH
solution\\ $\bar{u}^{\mathrm{NSH}}$&\;\;\;\;\;\;\;\;vertically integrated
radial velocity of the dust in the NSH solution\\ $u_{\mathrm{r}}$,
$u_{\varphi}$& radial and azimuthal velocity of the gas (solution of the
NSe)\\ $w_{\mathrm{r}}$, $w_{\varphi}$& radial and azimuthal velocity of the
dust (solution of the NSe)\\ $\bar{u}^{\mathrm{V}}$&vertically integrated
radial velocity of the dust (solution of the NSe)\\ $\Delta$&vertical lenght
scale in which dust and gas effect each other (NSe)\\ $\Delta_E$&lenght scale
of the classical Ekman layer\\ $\delta_{\mathrm{p}}$&drift induced by the
inclusion of viscosity\\ \hline
\end{tabular}
\caption{Variables used in the course of this paper.}
\end{center}
\end{table*}

\end{document}